# Effect of ECAP and heat treatment on mechanical properties, stress relaxation behavior and corrosion resistance of a 321-type austenitic steel with increased δ-ferrite content

V.N. Chuvil'deev[1], A.V. Nokhrin[1,(*)], N.A. Kozlova[1], M.K. Chegurov[1], V.I. Kopylov[1,2], M.Yu. Gryaznov[1], S.V. Shotin[1], N.V. Melekhin[1], C.V. Likhnitskii[1], N.Yu. Tabachkova[3,4]

[1] Lobachevsky State University of Nizhniy Novgorod, Nizhniy Novgorod, Russia

[2] Physical-Technical Institute, National Academy of Sciences of Belarus, Belarus, Minsk

[3] A.M. Prokhorov Institute of General Physics, RAS, Russia, Moscow

[4] National University of Science and Technology "MISIS", Russia, Moscow

nokhrin@nifti.unn.ru

**Abstract**

Hot rolled commercial metastable austenitic steel 0.8C-18Cr-10Ni-0.1Ti (Russian industrial name 08X18H10T, analog 321L) with strongly elongated thin δ-ferrite particles in its microstructure was the object of investigations. The lengths of these δ-particles were up to 500 μm, the thickness was 10 μm. The formation of the strain-induced martensite as well as the grinding of the austenite and of the δ-ferrite grains take place during ECAP. During the annealing of the UFG steel, the formation of the σ-phase particles takes place. These particles affect the grain boundary migration and the strength of the steel. However, a reduction of the Hall-Petch coefficient as compared to the coarse-grained (CG) steel due to the fragmentation of the δ-ferrite particles was observed. The samples of the UFG steel were found to have 2-3 times higher stress relaxation resistance as compared to the CG steel (a higher macroelasticity stress and a lower stress relaxation magnitude). The differences in the stress relaxation resistance of the UFG and CG steels were investigated. ECAP was shown to result in an increase in the corrosion rate and in an increased tendency to the intergranular corrosion (IGC). The reduction of the corrosion resistance of the UFG steel was found to originate from the increase in the fraction of the strain-induced martensite during ECAP.

**Keywords**: Austenitic steel, fine-grained microstructure, strength, relaxation resistance, corrosion resistance

(*) Corresponding author (nokhrin@nifti.unn.ru)

**Introduction**

Coarse-grained austenitic stainless steels Fe-Cr-Ni are used widely in nuclear power engineering, in oil and chemical industry. The austenitic steels are applied in fabricating the highly responsible products intended for operation in corrosion-aggressive ambient [1-8].

The problem of increasing the strength of the austenitic steels preserving their high resistance to the intergranular corrosion (IGC) is one of the key problems of the materials science [5-11]. This makes the traditional approach to increasing the strength consisting in annealing leading to the nucleation of the chromium carbide particles at the austenite grain boundaries inapplicable [12-16]. In this connection, engineers are developing novel methods of simultaneous increasing the strength and the corrosion resistance of the austenite steels.

Forming the ultrafine-grained (UFG) structure is one of the popular methods of improving the strength and the operational characteristics of stainless alloys [17-22]. At present, various methods of Severe Plastic Deformation (SPD) are applied to form the UFG structure – Equal Channel Angular Pressing (ECAP) [17-21], high pressure torsion [22-24], rotary swaging [25, 26], extrusion [26, 27] etc. In spite of certain success in the improvement of the hardness and strength of steels, it should be noted that the application of SPD leads to strain-induced decomposition of austenite often [17, 19, 21, 29-31]. It may affect the corrosion resistance of the UFG austenitic steels negatively. In this connection, the applied problem of choice of the optimal regimes of heat-deformation processing of austenitic steels, which allow increasing the strength of the ones without the reduction of the corrosion resistance is relevant.

A problem of providing a high stress relaxation resistance of the austenitic steels is even more complex. The problem of increasing the relaxation resistance is especially important in the development of machine-building hardware providing simultaneously high characteristics of fatigue, creep resistance, stress corrosion cracking resistance, etc. [32-36]. The high stress relaxation resistance determines the capability of the hardware to provide the necessary level of downforce during a long operation time [37, 38]. The improvement of the stress relaxation resistance of the materials with simultaneous proving a high strength will allow increasing the downforce of the hardware and keeping it during a notably longer operation time. Plenty of experimental and theoretical works was devoted to the problem of investigation of the stress relaxation resistance mechanisms for the coarse-grained (CG) materials [39-42]. For CG materials, it is supposed usually that the higher the level of internal stresses, the lower the stress relaxation depth (the magnitude of the decrease in the stress in given time interval). Therefore, the strain strengthening is a traditional method of increasing the stress relaxation resistance. From this viewpoint, the fine-grained metals and alloys fabricated using the SPD methods are promising candidates for application as the base materials for the heavy-duty relaxation-proof hardware.

The analysis of the literature shows that SPD may lead to an increase as well as in a decrease in the stress relaxation resistance of metals [43-52]. One should outline the works [43-47], where a faster and stronger reduction of the stresses in time in the UFG metals was demonstrated. Some authors related it to grain boundary sliding [43, 44, 48-51] or to interaction of the lattice dislocations with the grain boundaries [43, 46, 47, 50, 51], which may occur during the stress relaxation tests of the UFG materials along with accommodative redistribution of lattice dislocations.

The present work was aimed at studying the effect of SPD and annealing on the relaxation resistance and the resistance to IGC of the Russian metastable austenitic steels 0.8%C-18%Cr-10%Ni-0.1%Ti (Russian industrial name 08X18H10T, Russian analog of steel 321L). This steel is used widely in nuclear mechanical and power engineering for making the machine building hardware operated in the condition of simultaneous impact of elevated temperatures, mechanical loads, and corrosion-aggressive ambients. In particular, a low strength and high stress relaxation rate in the austenite steels result in difficulties in operations of assembling and disassembling the products after long-term operation. An increased content of δ-ferrite is a special feature of the object of investigations. It is a defect of casting or of heat treatment of the cast workpieces but is present in the bulk austenite steel often.

**Materials and methods**

The Russian commercial metastable austenitic steel 08X18H10T (composition: Fe-0.08wt.%C-17.9wt.%Cr-10.6wt.%Ni-0.5wt.%Si-0.1wt.%Ti) was the object of investigations. The formation of the UFG microstructure in the steel was performed by ECAP. The workpieces of 14×14×140 mm in sizes were cut out from hot-rolled rods of 20 mm in diameter. Prior to ECAP, the rods were annealed at 1050 ºC for 30 min followed by quenching in water. ECAP was performed using Ficep® HF400L press (Italy). The angle of crossing the working channel and the output one was $\pi/2$. In the ECAP regime used, the workpiece rotated at the angle of $\pi$ around its longitudinal axis during every cycle (regime "C", see [52]). The ECAP rate was 0.4 mm/s. The ECAP temperatures were 150 and 450 °C, the number of pressing cycles (N) varied from one to four.

The investigations of the steel microstructure were carried out using Jeol® JSM-6490 and Tescan® Vega™ Scanning Electron Microscopes (SEMs) and Jeol® JEM-2100F Transmission Electron Microscope (TEM). X-ray diffraction (XRD) phase analysis of the stainless steels was carried out using Shimadzu® XRD-7000 X-ray diffractometer ($CuK_\alpha$ emission, recording in the Bragg-Brentano scheme in the range of angles $2\theta = 30\text{-}80º$ with the scan rate 1 º/min). The crystal lattice parameters were determined and the mass fractions of the phases were calculated by Rietveld method.

The microhardness ($H_v$) of the steel was measured with Duramin® Struers™ 5 microhardness tester. The uncertainty of the microhardness measurements was ±50 MPa.

For the mechanical tests, flat double-blade shaped specimens were made by electric spark cutting. The sizes of the working part were 2×2×3 mm. The tension tests were carried out using Tinius Olsen® H25K-S machine with the strain rate $3.3 \cdot 10^{-3}$ $s^{-1}$ (the tension rate was $10^{-2}$ mm/s). The tension tests were performed at the room temperature (RT) and in the temperature range 450–900 °C. The specimens were heated up to the testing temperatures in 5 min. The specimens were kept at the testing temperatures for 10 min to establish the thermal equilibrium. In the course of the tests, the curves stress ($\sigma$) – strain ($\varepsilon$) were recorded. From these curves $\sigma(\varepsilon)$, the magnitudes of the ultimate strength ($\sigma_b$) and of the maximum relative elongation to failure ($\delta$) were determined.

The fractographic analysis of the fractures after the tension tests was carried out using Jeol® JSM-6490 SEM. The macrostructure of the specimens after the failure tests was investigated using Leica® IM DRM metallographic optical microscope. The investigations of microstructure and the microhardness measurements were performed in the fracture zones ("deformed area") and in the non-deformed areas near the capturers.

The stress relaxation tests were performed according the technique described in Appendix A to the paper [53]. For the tests, the rectangular specimens of 3×3 mm in cross-sections and of 6 mm in height were made. The specimens were loaded with the rate 0.13%/s during 0.3 s. Afterwards, the specimens were kept under a constant stress ($\sigma_i$) during given stress relaxation time ($t_r = 60$ s). In the course of stress relaxation, a curve of the stress on the testing time $\sigma_i(t)$ was acquired. Afterwards, the next loading step was performed. As a result of experiment, a dependence of the stress relaxation magnitude $\Delta\sigma_i$ on the magnitude of the summary load applied $\Delta\sigma_i(\sigma)$ was obtained. The dependence obtained was used also to determine the macroelasticity stress ($\sigma_0$) and the yield strength ($\sigma_y$).

The resistance of the steels to the intergranular corrosion (IGC) was investigated using R-8 potentiostat-galvanostat according to Russian National Standard GOST 9.914-91 by double loop electrochemical potentiokinetic reactivation (DLEPR) method. The DLEPR tests were conducted at RT in an aqueous solution 10%$H_2SO_4$ +0.0025 g/l KSCN. An auxiliary electrode was made from a Pt grid, the reference electrode – from chlorine silver, the investigated specimen served as the working electrode. The investigated specimen was cathode polarized at the potential $\varphi = -550$ mV for 2 min. The curves voltage – current density were recorded in the range of potentials from –550 mV to +1200 mV with a rate of 3 mV/s. The tendency of the steel to IGC was determined from the ratio of the areas under the passivation curve ($S_1$) and under the reactivation one ($S_2$): $K = S_1/S_2$. According to GOST 9.914-91, the increasing of the coefficient K up to K = 0.11 means that the austenite steel demonstrates an increased tendency to IGC.

The Tafel curves ln(i) – E were measured in the same medium. From the Tafel curves, the corrosion current densities ($i_{corr}$, mA/cm$^2$) and the corrosion potentials ($E_{corr}$, mV) were obtained by standard method. Prior to the corrosion investigations, the surfaces of specimens of 5×10×10 mm in size were subjected to mechanical grinding and polishing. From the results of measuring the $i_{corr}$, the corrosion rate was calculated using the formula: $V_{corr} = 8.76 i_{corr} M / \rho F$, where $\rho$ is the density of iron [g/cm$^3$], M – molar mass [g/mol], F = 96500 C – Faraday's constant. The verification tests of resistance against IGC were conducted according to GOST 6232-2003 by the boiling of the specimens in a solution of 25% $H_2SO_4$ + $CuSO_4$. The character of the surface destruction after the corrosion tests was analyzed using Leica® IM DRM metallographic optical microscope.

To study the thermal stability of the structure and properties of the UFG steel, the specimens were annealed in air in the temperature range from 100 up to 900 °C. The isothermic holding time was 60 min. Th uncertainty of maintaining the temperature was ± 10 °C. The specimens were cooled down in water.

**Results**

Microstructure investigations

As initial state, stainless steel had a uniform austenite microstructure (Figs. 1a-1d). The mean austenite grain sizes were ~20 μm. The thin (up to 10 μm in thickness) strips of the ferrite δ-phase elongated along the deformation direction were observed in the microstructure of the CG steel (Figs. 1a-1d). The lengths of the δ-ferrite stripes were ~500 μm. The lattice dislocations (Fig. 1f) as wee as few micron- and submicron-sized titanium carbide and carbonitride particles (Fig. 1e) were observed inside the austenite grains.

After the first ECAP cycle, the macrostructure of the steel workpieces comprised of alternating macro-bands of localized strain (Fig. 2). After N = 4 ECAP cycles, the specimens had a uniform macrostructure.

Fig. 3a presents the XRD curves from the steel specimens in the initial state and after ECAP. An XRD peak 111 α(δ)-phase (PDF 00-006-0696) is seen clearly in the XRD curve of the CG steel at the diffraction angel 2θ ~ 45º near the highly intensive XRD peak 110 γ-Fe (PDF 01-071-4649). The results of the XRD phase analysis evidence the mean mass fraction of the δ-phase in the steel in the initial state to be ~1.5–3 %. The lattice parameter of the δ-phase in the steel Fe-Cr-Ni-Ti was 2.8869 Å, the one of the γ-phase was 3.5875 Å.

ECAP leads to an increase in the fraction of the α-phase because of appearing the strain-induced martensite. The scale and dynamics of the increasing of the strain-induced martensite with increasing the number of cycles depends on the ECAP temperature (Fib. 3b). After ECAP at 150 °C, the mass fraction of the α(δ)-phase was 5.9–7.7% and didn't change essentially with increasing number of ECAP cycles up to N = 4. The increasing of the SPD temperature up to 450°C resulted in more intensive decomposition of the γ-phase (austenite) – the mass fraction of the α(δ)-phase increased from 3.6% after N = 1 cycle up to 17.4% after N = 4 ECAP cycles. Note that the effect of strain-induced decomposition of the γ-phase during ECAP of the austenitic steel is known for a long enough time [19-23]. At the same time, it is worth noting that the effect of accelerated strain-induced decomposition of austenite at higher SPD temperatures is an unexpected ones (see [24]).

ECAP resulted in a decreasing of the intensity and in the broadening of the XRD peaks from the α- and γ-phases. The half width at half maximum (HWHM) of the 111 XRD peaks of α(δ)-Fe and of the (110) ones of γ-Fe for the coarse-grained steel were 0.196º and 0.193º, respectively. In the UFG steel after N=4 ECAP cycles at 150 ºC, the HWHMs of the 111 α(δ)-Fe and (110) γ-Fe XRD peaks were 0.300º and 0.277º, respectively while for the UFG steel specimens after N=4 ECAP cycles at 450 ºC – 0.407º and 0.289º, respectively. The lattice constants of α(δ)-Fe and γ-Fe for the UFG steel after N = 4 ECAP cycles ($a_α$ = 2.8718 Å, $a_γ$ = 3.5863 Å – $T_{ECAP}$ = 150 ºC; $a_α$ = 2.8780 Å, $a_γ$ = 3.5897 Å – $T_{ECAP}$ = 450 ºC) were close to the ones of ferrite and austenite in the coarse-grained steel. This allow suggesting small grain sizes (small sizes of the coherent scattering regions) to introduce the major contributions in the broadening of the XRD peaks after N=4 ECAP cycles.

Along with the austenite strain-induced decomposition during ECAP, a grinding of the steel grain microstructure was observed. After N = 4 ECAP cycles at 150 and 450 °C, an UFG microstructure with mean grain sizes of 0.3 and 0.5 μm, respectively formed in steel (Fig. 4). For the specimens of steel obtained by ECAP at 150 °C, the crossing localization micro-bands, which lead to different orientation of the austenite grains were observed at the microscopic level (Fig. 4a, c, d). The microstructure of the specimens after ECAP at 450 °C was more uniform, no shear microbands manifested clearly were observed (Fig. 4d). The nanotwins are seen in some austenite grains (Fig. 4d), which can be classified as the strain-induced martensite according to [17-20]. No nucleation of the chromium carbide particles was observed in the steel microstructure after ECAP. No δ-phase particles were reveled in the UFG microstructure during the metallographic and SEM investigations that allows making a conclusion on a strong fragmentation of these ones during ECAP. The presence of separate point reflections in the electron diffraction pattern evidenced the presence of the high-angle grain boundaries in the UFG steel obtained by ECAP at 450 °C (Fig. 4f). The electron diffraction patterns from the specimens of the UFG steel after ECAP at 150 °C were more blurry (Fig. 4b).

The investigations of the thermal stability of the UFG microstructure during annealing demonstrated the temperature of recrystallization in the UFG steel (N = 4, $T_{ECAP}$ = 450 °C) to be $T_1$ = 750 °C. The recrystallization had a clearly expressed abnormal character accompanied by a formation of a multi-grained structure. After annealing at 750 °C, 1 hour, the recrystallized metal regions with the mean grain sizes of 5-7 μm were observed in the microstructure of the UFG steel. The volume fraction of these regions was ~3% or less. At increased annealing temperatures, an increase in the volume fraction of the recrystallization metal as well as an increase in the mean grain sizes were observed – after annealing at 900 °C, 1 hour, an equiaxial austenite microstructure with the mean grain sizes of 8-12 μm formed in the steel (Fig. 5). Increasing the number of ECAP cycles up to N = 4 at $T_{ECAP}$ = 450 °C didn't result in any change of the recrystallization temperature $T_1$ but was accompanied by a decrease in the mean recrystallized grain sizes (Fig. 5).

The XRD phase analysis demonstrated a decrease in the mass fraction of the α(δ)-phase with increasing annealing temperature up to 600°C (1 h). After annealing at 800 °C, the mass fraction of the α(δ)-phase was beyond the measurement uncertainty ±1 wt.% (the intensity of the XRD peaks from the α(δ)-phase didn't exceed the noise level) regardless to the ECAP regimes.

The *in situ* TEM investigations demonstrated the nucleation of the light-colored nanoparticles in the UFG steel when heating up to 600 °C. The mean size and the volume fraction of the particles increased with increasing heating temperature. After heating up to 800 °C and holding for 0.5 hrs, the mean particle size was about 50 nm (Figs. 6, 7). Because of the presence of the α(δ)-phase having an essential residual magnetization in the steel, we failed to analyze the composition of the nucleated second phase nanoparticles by EDS. We suggest these particles to be the σ-phase ones. The peaks from the σ-phase were absent in the XRD curves from the annealed specimens, probably, due to small sizes of the nucleated particles. Earlier, the possibility of nucleation of the σ-phase particles during annealing of UFG steel 08X18H10T was reported in [18, 19].

Mechanical properties at room temperature

As shipped, the CG steel had the macroelasticity stress ($σ_o$) and the yield strength ($σ_y$) equal to 205 MPa and 380 MPa, respectively. The processing by ECAP resulted in an improvement of the mechanical properties of the steel. The $σ_o$ increased up to 340 MPa and 425 MPa and the $σ_y$ – up to 940 and 1070 MPa, respectively with increasing number of ECAP cycles from N = 2 to 4 at $T_{ECAP}$ = 450 °C (Fig. 8a). The magnitudes of the yield strength and of the macroelasticity stress of the UFG steel depended on the ECAP temperature weakly – the $σ_y$ increased from 1070 to 1145 MPa and the $σ_o$ decreased from 425 to 410 MPa with decreasing ECAP temperature from 450 down to 150 °C (N = 4).

The analysis of the yield strength dependence on the mean grain sizes has shown that this dependence can be interpolated by a straight line in the $\sigma_y - d^{-1/2}$ axes with a good precision (Fig. 8b). This evidences the Hall-Petch relation to hold:

$$\sigma_y = \sigma_o + K \cdot d^{-1/2}, \qquad (1)$$

where K is the grain boundary hardening coefficient (Hall-Petch coefficient) describing the contribution of the grain boundary structure in the strength of the metal. The mean value of the coefficient K determined from the dependence in Fig. 8b is $K = 0.46$ MPa·m$^{1/2}$.

Fig. 9 presents the dependencies of the macroelasticity stress (Fig. 9a) and of the yield strength (Fig. 9b) on the temperature of 1 hour annealing of the UFG steel. One can see in Fig. 9a that the dependencies $\sigma_o(T)$ have a three-stage character. The first stage (RT–300 °C) is characterized by a constant value $\sigma_o$. At the second stage of annealing (500–600 °C), an increase in the macroelasticity stress was observed, which probably originates from the nucleation of the second phase particles (see above). At the third stage of annealing, at the temperatures greater than 600 °C, a decrease of $\sigma_o$ down to the values typical for the CG steel in the as–shipped state was observed. The softening of the UFG steel at this stage of annealing was related to the recrystallization leading to an increase in the grain sizes.

The dependence of the yield strength on the annealing temperature had usual two-stage character (Fig. 9b). Note that the increase in the macroelasticity stress at 600 °C originating from the nucleation of the second phase particles didn't lead to the increase in the yield strength as one could expect because of the Hall-Petch law (see (1)). This result evidences active grain boundary recovery processes to go at this stage. These processes lead to a decrease in the defect density in the grain boundaries [50].

Note also that the values of the macroelasticity stress and of the yield strength of the CG steel almost didn't change when annealing at the temperatures up to 700 °C. After heating up to higher temperatures, an insufficient decrease in $\sigma_o$ and in $\sigma_y$ was observed. After annealing at 900 °C, the values of the macroelasticity stress and of the yield strength for the CG steel and for the

UFG one were close to each other.

The stress–strain tension curves $\sigma(\varepsilon)$ of the CG steel specimens and of the UFG ones at RT are presented in Fig. 10. The $\sigma(\varepsilon)$ curve of the CG steel had a classical form, with a long strain hardening stage. The magnitude of the tensile strength of the CG steel was $\sigma_b = 720$ MPa. This is a very high value, which probably originates from the presence of the δ–ferrite particles and from relatively small grain sizes in austenite (~20 μm) in the hot-deformed steel. In the stress–strain tension curves $\sigma(\varepsilon)$ of the UFG steel specimens, there are a short stage of stable strain flow, which transforms into a stage of localization of the strain. Increasing the ECAP temperature from 150 up to 450 °C resulted in an insufficient increase in the duration of the uniform strain stage.

The tension tests of the CG and UFG steel specimens demonstrated the formation of the UFG structure by ECAP (N = 4 at T = 150 °C) to result in a decreasing of the elongation to failure (δ) of the steel from 125 down to 45% and in an increasing of the ultimate strength from 720 up to 1100 MPa (Table 1). ECAP at higher temperature (450 °C) resulted in an insufficient decreasing of the ultimate strength down to 1020 MPa and in an increasing of the elongation up to δ ~ 60%.

The fractographic analysis revealed three characteristic zones on the fractures of the CG steel specimens and of the UFG ones after the tension tests. These are: the fibrous zone, the radial one, and the break (cut) zone (Fig. 11). It is worth noting that the cut zones in the CG steel occupied ~50 % of the whole fracture area. In the UFG steel after ECAP (N = 4), the cut zones occupied ~70%. So far, the formation of the UFG microstructure resulted in an increasing of the cut zone fraction of the fracture area and, hence, in a decreasing of the viscous component of the fracture.

Stress relaxation resistance

Fig. 12a presents the stress relaxation curves $\Delta\sigma_i(\sigma)$ for the specimens of the CG steel and of the UFG ones. The stress relaxation curve $\Delta\sigma_i(\sigma)$ for the CG steel had a classical three-stage character: one can distinguish the macroelasticity strain stage, the microplastic strain stage, and the macroplastic strain one clearly enough. Note that one can see the stress relaxation curves $\Delta\sigma_i(\sigma)$ to be close to each other at the stresses < 150-170 MPa; no essential differences in the stress relaxation magnitudes were observed. In the microplastic strain range (where the stress increased from 150-170 MPa up to 300-320 MPa), the stress relaxation magnitude $\Delta\sigma_i$ of the CG steel specimens began to increase drastically and reached ~15 MPa at the stress of 320 MPa. At further increasing of the stress up to 580-600 MPa (in the macroplastic strain region), a smooth increasing of the stress relaxation magnitude up to $\Delta\sigma_i$ ~20 MPa was observed.

The stress relaxation curves for the UFG steel specimens had more smooth character than the curves $\Delta\sigma_i(\sigma)$ for the CG steel ones. Note that the fairly expressed macroplastic strain stage was almost absent – as one can see in Fig. 12a, the microplastic strain stage transforms into the macroplastic strain one smoothly enough. The increasing of the of the number of the ECAP cycles resulted in a displacement of the curves $\Delta\sigma_i(\sigma)$ towards the higher stresses. One can see in Fig. 12a that the stress relaxation magnitude $\Delta\sigma_i$ ~20 MPa in the UFG steels subjected to N = 1 ECAP cycle was achieved at the stress of 670-690 MPa whereas in the UFG steels obtained by N = 4 cycles of processing by ECAP, this stress relaxation magnitude was achieved at the stress of 935–950 MPa ($T_{ECAP}$ = 450 ºC) and 990–1010 MPa ($T_{ECAP}$ = 150 ºC).

So far, one can conclude the processing of the austenitic steel by ECAP to result in the increasing of its stress relaxation resistance – in the increasing of the macroelasticity stress $\sigma_o$ (see above) and in the decreasing of the stress relaxation magnitude $\Delta\sigma_i$ at increased loads.

The recrystallization annealing resulted in a decreasing of the stress relaxation resistance parameters of the UFG steels – as one can see in Fig. 12b, the increasing of the annealing temperature above 650-700 ºC resulted in a displacement of the stress relaxation curves $\Delta\sigma_i(\sigma)$ towards the smaller stresses. After annealing at 800-900 ºC, the stress relaxation curves of the deformed steel specimens had usual tree-stage character corresponding to the stress relaxation curve $\Delta\sigma_i(\sigma)$ of the coarse-grained steel specimens (Fig. 12a).

Tension testing at elevated temperatures

Table 2 presents the dependencies of the ultimate strength and of the elongation to failure on the testing temperature for the coarse-gained steel specimens and for the UFG ones obtained in different ECAP temperatures. Fig. 13 presents the stress–strain curves $\sigma(\varepsilon)$ for the tension tests at elevated temperatures.

The stress–strain curves $\sigma(\varepsilon)$ for the CG steel specimens had the form typical for high-plasticity materials (Fig. 13a). The duration of the localized plastic strain stage was much smaller than of the uniform strain one. The curves $\sigma(\varepsilon)$ for the UFG steel specimens at the testing temperatures of 750 and 800 ºC had the form typical for highly plasticity materials – the stage of an insufficient strain hardening transformed into a long state of stable strain flow (Figs. 13b, c). The analysis of the dependencies presented in Fig. 13 demonstrated the increasing of the temperature from RT up to 750 °C to result in a monotonous decreasing of the ultimate strength (the flow stress) from 720 MPa down to 250 MPa for the CG steel and from 950–1100 MPa down to 240–290 MPa for the UFG steel, respectively.

Note that the increasing of the testing temperature resulted in a nonmonotonous variation of the elongation to failure for the UFG steel that differs from the same dependencies for the CG steel. The analysis of the data presented in Fig. 13b shows the elongation to failure for the CG steel to decrease monotonously from 125% down to 70% with increasing testing temperature from RT up to 750 °C. The character of the dependence $\delta(T)$ for the UFG steel was more complex – the elongation to failure decreased insufficiently with increasing testing temperature from RT up to 450 °C. At further increasing of the temperature from 450 °C up to 750–800 °C, the elongation to failure of the UFG steel increased and was several times higher than the $\delta$ of the CG steel. For the UFG steel specimens obtained by ECAP at $T_{ECAP} = 150$ °C, the elongation to failure at the testing temperature of 750 °C reached 250 %. Further increasing of the testing temperature resulted in a decreasing of the elongating to failure for the UFG steel specimens again.

The fractographic analysis of the fractures (Fig. 14) demonstrated the areas of the fibrous zones and of the radial ones to increase and the areas of the cut zones – to decrease with increasing testing temperature. At the testing temperature of 600 °C, the cur zone area didn't exceed 5–10% of the whole fracture area. For the UFG steels ($T_{ECAP}$ = 450 °C), the cut zones were absent that also evidences an increased plasticity of the UFG material as compared to the coarse-grained state.

In our opinion, the non-monotonous character of the dependence of the elongation to failure on the testing temperature for the UFG steel originates from the recrystallization processes in the UFG steel starting after annealing at ~650–700 °C.

In particular, the microhardness testing results (Table 2) evidence the intensive recrystallization at the high-temperature strain of the UFG steel. The microhardness measurements of the specimens after the tension tests demonstrated the increasing of the testing temperature from 450 up to 800 °C to result in a decreasing of the microhardness both in the deformed areas and in the non-deformed ones. This conclusion is supported by the results of the microstructure investigations in the deformed areas of the UFG steel specimens and in the non-deformed ones after the tension tests (Fig. 15). As one can see from Fig. 15 and from the data presented in Table 3, the testing at 800 °C resulted in the formation of a well uniform fine-grained structure. No essential grain growth was observed. The mean grain sizes in the deformed parts were slightly smaller than in the non-deformed ones.

Corrosion resistance

Fig. 16a presents the Tafel curves ln(i)–E for the coarse-grained steel specimens and for the UFG ones. The results of the electrochemical testing are summarized in Table 3. The curves ln(i)–E had usual character. One can see the coarse-grained steel specimens to have smaller corrosion rates than the UFG ones. For the UFG steel specimens obtained by ECAP at T = 450 °C, the values of mean corrosion current density $i_{corr}$ (of the mean corrosion rate $V_{corr}$) were 10–15% greater than the same characteristics for the UFG steel specimens obtained by ECAP at 150 °C.

Fig. 16b presents the curves i(E) illustrating the results of testing by the DLEPR method according to GOST 9.914-91. The results of these tests are summarized in Table 3. It follows from the data presented in Table 3 that the ratios of the areas under the passivation curves ($S_1$) and the reactivation ones ($S_2$) (K = $S_1/S_2$) were small and appeared to be much less than the ultimate value $K_{max}$ = 0.11. This result evidences both coarse–grained steel and UFG one to be highly resistant against IGC. At the same time, the magnitudes of the coefficient K for the UFG steel specimens were 1.5–2.5 times higher than for the CG ones. The metallographic analysis of the surfaces has shown the large elongated δ-ferrite particles to be the places of accelerated corrosion destruction of the surfaces in the DLEPR testing (Fig. 17a). No IGC traces were observed on the surfaces of the UFG steel specimens (Fig. 17b).

The results of standard tests of the resistance against IGC according to GOST 6232-2003 confirmed high corrosion resistance of the UFG steels. As one can see in Fig. 18a, after testing during 24 hrs, the corroded elongated δ-ferrite particles were observed on the CG steel surfaces. In some areas of the surfaces, the IGC corrosion defects or the pitting corrosion no more than 10–15 μm in depth are seen. On the surfaces of the steel specimens with the UFG structure formed as a result of 1 or 2 ECAP cycles, few corrosion pits were observed. On the surfaces of the UFG steel specimens (N = 3, 4), the corrosion defects were absent (Fig. 18b).

So far, the UFG steel specimens have high strength, stress relaxation resistance, and high resistance against the intergranular corrosion simultaneously. It allows an efficient application of the UFG steel for making the stress relaxation–proof machine–building hardware utilized in the conditions of enhanced loads and corrosion–aggressive media.

**Discussion**

<u>Investigation of thermal stability</u>

First, one should pay attention to the fact of nucleation of the second phase particles during the annealing of the UFG austenitic stainless steel. It is interesting to note that the nucleation of the second phase particles was observed not in all grains. In our opinion, the nucleation of the particles goes preferentially inside the grains of α–phase, the lattice constant of which is much less than the one of the γ–phase. It leads to a formation of a strongly supersaturated solid solution (of chromium) in the α–phase grains and, as a consequence, to its nucleation at further heating up. This assumption allows suggesting the nucleation of ferromagnetic σ–phase particles Fe-Cr to take place in the course of heating up. We suppose the nucleation of chromium carbide $Cr_{23}C_6$ particles to be hardly probable in this case since steel contains titanium, which reacts chemically with carbon and forms titanium carbide TiC [10, 18, 20, 22]. The possible nucleation of the σ-phase particles during annealing of metastable UFG austenitic steel Fe-Cr-Ni-Ti was reported in [18, 19].

The analysis of the grain growth process revealed the grain growth activation energy ($Q_R$) determined from the slope of the dependence $\ln(d^n-d_0^n) - T_m/T$ to be 6.0–8.3 $kT_m$ (~ 90–125 kJ/mol) (Fig. 5). The uncertainty of determining the activation energy $Q_R$ was ±1 $kT_m$. The calculated activation energy depends on the number of ECAP cycles or on the ECAP temperature weakly. In the calculations, the magnitude of coefficient $n$ was taken to be $n = 4$ that corresponds to the case of the migration of grain boundary with the particles nucleated at the ones [52]. The melting point of steel was taken to be $T_m$ = 1810 K. Note that the recrystallization activation energy was ~20–30% smaller than the equilibrium activation energy of the grain boundary diffusion in austenite $Q_b$ ~ 10.6 $kT_m$ (159 kJ/mol [52]). In our opinion, this result evidences the nonequilibrium grain boundaries in the UFG steel obtained by ECAP to contain an increased concentration of defects – the orientation mismatch dislocations (OMDs) and the products of the delocalization of the ones (the tangential components of Burgers vectors of the delocalized dislocations) [52]. The increased density of defects in the grain boundaries leads to an increasing of the free volume of the grain boundaries in the UFG material [52] and, as a consequence, to a decreasing of the activation energy of the grain boundary diffusion [52].

Note also that at $n = 2$, the recrystallization activation energy $Q_R$ takes non-physical values (3–4.3 $kT_m$ ~ 45–63 kJ/mol), which appear to be smaller than the activation energy of diffusion in the iron melt (see [52]). In our opinion, it evidences indirectly the nucleating nanoparticles to affect the grain boundary migration in the deformed austenite steel essentially.

Mechanical properties

The yield strength in the austenitic steel can be calculated using Hall-Petch equation (1) where the magnitude of the macroelasticity stress in the first approximation can be calculated as the sum of the following contributions:

$$\sigma_0 = \sigma_{PN} + \sum A_i C_i + \alpha_1 MGb\sqrt{\rho_v} + 2\alpha_2 MGb/\lambda, \qquad (2)$$

where $\sigma_{PN}$ is the stress of resistance of the crystal lattice (the Peierls-Nabarro stress), $\sigma_c = \sum A_i C_i$ accounts for the contributions of the doping elements into the strengthening of austenite ($A_i$ is the contribution of the $i^{th}$ doping element, the concentration of which is $C_i$ into the hardening of austenite) [55], $\sigma_d = \alpha_1 MGb\sqrt{\rho_v}$ is the contribution of the hardening dislocations ($\rho_v$ being the density of the lattice dislocations) [55], $\sigma_p = 2\alpha_2 MGb/\lambda$ is the contribution of the second phase particles ($\lambda$ is the distance between the particles), where G = 81 GPa is the shear modulus, b = 0.258 nm is the Burgers vector, $\alpha_1$ = 0.3-0.67 is a numerical coefficient depending on the character of the distribution and of the interaction of the lattice dislocations, $\alpha_2$ = 0.5 is a numerical coefficient, M = 3.1 is the orientation factor (the Taylor coefficient).

According to [55, 56], the contribution of the crystal lattice of the doped austenite for steel and the heat-resistance nickel alloys is $\sigma_{PN}$ = 60-70 MPa. The contribution of the second phase particles can be neglected in the first approximation since the nucleated particles were large enough (Fig. 1d) and were located far enough from each other: at $\lambda$ = 5-10 μm, the contribution of the second phase particles is $\sigma_d \sim$ 10 MPa.

Since the contribution of Ni in the hardening of austenite is small [55], one can suggest the dislocation hardening makes the major contribution in the magnitude of the macroelasticity stress of the austenite steel ($\sigma_0$ = 240 MPa). The magnitude of $\sigma_d = \sigma_0 - \sigma_{PN}$ = 170–180 MPa at $\alpha_1$ = 0.3 corresponds to the density of lattice dislocations of $\rho_v \sim 8 \cdot 10^{13}$ m$^{-2}$ whereas at $\alpha_1 = 0.67$ to $\rho_v \sim 1.5 \cdot 10^{13}$ m$^{-2}$. This estimate of $\rho_v$ matches well to the data reported in the literature [17-20].

For the mean value of $K = 0.46$ MPa·m$^{1/2}$ (see above) and $d_\gamma \sim 20$ μm, the contribution of the grain boundary hardening $\sigma_{gb} = K \cdot d^{-1/2}$ in the CG austenitic steel is ~105 MPa.

The calculated value of the yield strength of the CG austenitic steel $\sigma_y = 240$ MPa + 105 MPa = 245 MPa was lower than the experimentally measured value (380 MPa).

In our opinion, there are two main origins of the discrepancy between the results of calculations and the experimental data.

First, it is worth noting the large particles of δ-ferrite in the microstructure of austenitic steel, which can impede the micro- and macroplastic strain. Traditional approach to the calculation of the yield strength of the steel with such a composite structure consists in accounting for the volume fraction and the yield strength of δ-ferrite: $\sigma_y = f_\gamma \sigma_{y(\gamma)} + f_\delta \sigma_{y(\delta)}$ where $f_\gamma$ and $f_\delta$ are the volume fractions of austenite (γ-Fe) and of δ-ferrite, $\sigma_{y(\alpha)}$ and $\sigma_{y(\delta)}$ are the yield strength of austenite and δ-ferrite, respectively. Unfortunately, at present it is impossible to measure the yield strength of δ-ferrite $\sigma_{y(\delta)}$ correctly. In this connection, it is impossible to estimate the effect of such meso-barriers on the ultimate strength correctly at present.

The second origin is, from our viewpoint, is the effect of the structural and phase state of the grain boundary on the magnitude of the Hall-Petch coefficient K. It leads to an essential difference of the mean value of K calculated from the dependence $\sigma_y - d^{-1/2}$ from the Hall-Petch coefficients in the coarse–grained steel ($K_0$) and in the UFG steel ($K_1$).

Note also that the intensities of increasing of the macroelasticity stress $\sigma_o$ and of the yield strength $\sigma_y$ with increasing number of ECAP cycles (N) were different (Fig. 9a). Analysis of the data presented in Fig. 9a shows the magnitude of $\sigma_{gb} = \sigma_y - \sigma_o = Kd^{-1/2}$ in the initial state to be 175 MPa and to increase up to $\sigma_{gb} = 645$-$655$ MPa with increasing N up to 3–4 ($T_{ECAP} = 450$ °C). Note that at the same time, the magnitude of the grain boundary hardening coefficient K calculated according to the formula K = ($\sigma_y$ - $\sigma_o$)d$^{1/2}$ (see (1)) decreased monotonously with increasing number of ECAP cycles. The magnitude of the Hall-Petch coefficient K for the CG steel was 0.78 MPa·m$^{1/2}$. After N = 3 and N = 4 ECAP cycles at 450 °C, it decreased down to 0.46 and 0.35 MPa$^{1/2}$, respectively. Similar effect was observed for the UFG steel specimens obtained by ECAP at $T_{ECAP}$ = 150 °C.

In our opinion, the decreasing of the coefficient K in ECAP is related to the fragmentation of strongly elongated δ-ferrite particles (up to 10 μm in thickness and up to 500 μm long). The harder δ-ferrite particles crossing the austenite grains often (Fig. 1) can impede the propagation of the strain in the austenite grains as well as the "transfer" of the strain from one austenite grain to another. In our opinion, strong fragmentation of the harder particles during ECAP helps eliminating the additional type of the "barrier" obstacles and promotes the strain at the micro- and macrolevels. In our opinion, the fragmentation of the large elongated δ-ferrite particles is one of the possible origins of the presence of the uniform strain flow stage in the stress–strain tension curves at room temperature (Fig. 10).

Taking for the CG steel K = 0.78 MPa·m$^{1/2}$ (see above) and $d_\gamma$ ~ 20 μm, one gets the contribution of the grain boundary hardening in the CG steel $\sigma_{gb}$ ~ 175 MPa. In this case, one gets the value of yield strength of the CG austenite steel calculated taking into account the correction for the magnitude of K $\sigma_y$ = 240 MPa + 175 MPa = 415 MPa. The calculated value of the yield strength matches well to the one measured experimentally ($\sigma_y$ = 380 MPa).

The magnitudes of the macroelasticity stress and of the yield strength for the UFG steel after N = 4 ECAP cycles were 410–425 MPa and 1070–1145 MPa, respectively.

Since the contributions $\sigma_{PN}$, $\sigma_c$, and $\sigma_p$ don't change doting ECAP, in our opinion, the increasing of the macroelasticity stress in 30-45 MPa originated from the increase in the density of the lattice dislocations up to ~1.2·10$^{14}$ m$^{-2}$ (at $\alpha_1$ = 0.3) whereas the increasing of the yield strength – from the decrease in the grain sizes down to the submicron level.

Stress relaxation resistance

As it has been shown above, the UFG steel has a higher stress relaxation resistance – the stress relaxation magnitude $\Delta\sigma_i$ in the UFG steel were much smaller at the same stress applied (Fig. 12a). Let us analyze the stress relaxation mechanisms in UFG steel underlying its improved stress relaxation resistance.

In general, the accommodative reconstruction of the defect structure (first of all – of the dislocation one) is well known to be the primary stress relaxation mechanism. In the coarse-grained materials at RT, the lattice dislocation glide in the field of the point defects distributed uniformly is such a mechanism most often. The dependence of the strain rate $\dot{\varepsilon}$ on the stress $\sigma$ in this case can be described by the following formula

$$\dot{\varepsilon} = \dot{\varepsilon}_0 \exp\left(-\frac{\Delta F}{kT}\left\{1 - \frac{\sigma}{\sigma^*}\right\}\right), \qquad (3)$$

where $\dot{\varepsilon}_0$ is the pre-exponential factor, $\Delta F$ is the activation energy of dislocation glide depending on the obstacles type, k is the Boltzmann constant, T is the testing temperature, and $\sigma^*$ is the non-thermal flow stress, which can be taken to be equal to the ultimate strength [55].

In the first approximation, the strain rate in the stress relaxation tests can be accepted to be proportional to the stress relaxation one: $\dot{\varepsilon} = \dot{\sigma}/E$, where E is the elastic modulus. The stress relaxation rate can be calculated as $\dot{\sigma} = \Delta\sigma/t_r$. Since the stress relaxation time $t_r = 60$ s and $E = 217$ GPa were the same for all specimens, the magnitude of the activation energy $\Delta F/kT$ can be determined from the slope of the dependence $\ln(\Delta\sigma) - 1-\sigma/\sigma_b$ (Fig. 19a).

As one can see in Fig. 19a, the dependence $\ln(\Delta\sigma) - 1-\sigma/\sigma_y$ for the CG steel has a two-stage character. The activation energy of dislocation glide in the microplastic strain range is $\Delta F_1 \sim 4.8$ kT (~0.62 Gb$^3$) that matches well to the data published in the literature (~ 0.5 Gb$^3$ for the steels AISI 304 and AISI 316 [57]). It allows concluding the gliding of lattice dislocations in the long-range stress field from other lattice dislocations to be main stress relaxation mechanism within the microplastic strain stage. At increased stresses, the activation energy of overcoming the obstacles tends to $\Delta F_2 \sim 0.9$ kT (~0.12 Gb$^3$). According to the classification of [57], the obstacles with $\Delta F <$

0.2 Gb$^3$ are the classified as the "weak" ones for the dislocation motion. In the case of the coarse-grained steel deformed in the macroscopic strain range, obviously, the austenite grain boundaries can be such obstacles.

In the case of the UFG steel, the stage with the increased $\Delta F_1$ ~4.9–6.2 kT (~0.63–0.80 Gb$^3$) was observed at small stresses only. In the range of micro- and macroplastic strain, the magnitude of activation energy of overcoming the obstacles was $\Delta F_2$ ~2.2–2.3 kT (~0.28–0.30 Gb$^3$). In the UFG metals, the grain boundaries are the main type of the obstacles for the gliding of the lattice dislocations. In this connection, one can suggest that the long microplastic strain stage characterizes the overcoming of the grain boundaries by the lattice dislocations.

Note that the magnitude of $\Delta F_2$ in the UFG steel (~0.28–0.30 Gb$^3$) is considerably greater than the one in the coarse-grained steel (~0.12 Gb$^3$).

The nonequilibrium grain boundaries in the UFG metals are known to contain an increased density of the OMDs featured by the density $\rho_b \Delta b$ and of the products of delocalization of the ones – the tangential ("sliding") components of the Burgers vectors of the delocalized dislocations and of the normal ones featured by the densities $w_t$ and $w_n$, respectively [52]. The defects introduced into the grain boundaries during ECAP generate the long–range internal stress fields, which impede the sliding of the lattice dislocations inside the austenite grains and prevent the formation of the dislocation clusters at the grain boundaries [52]. In our opinion, this factor is the primary origin of the increasing of the activation energy for overcoming the grain boundaries by the lattice dislocations $\Delta F_2$ in the UFG steel. This assumption is supported indirectly by the change of the actiation energy $\Delta F_2$ at the annealing of the UFG steel (Fig. 19b). As one can see in Fig. 12b, the recrystallization annealing of the UFG steel leads to the change in the character of the stress relaxation curves $\Delta\sigma(\sigma)$. The annealing of the UFG steel at the temperatures below 700 ºC (corresponding to the start of recrystallization) doesn't lead to any essential change of $\Delta F_2$ ~ 2.70–2.92 kT (0.34–0.37 Gb$^3$). After annealing at 750-800 ºC, the form of the $\ln(\Delta\sigma)$ – 1-$\sigma/\sigma_b$ dependence turned into a two-stage one while the magnitude of $\Delta F_2$ decreases monotonously from

1.39–1.49 kT (0.17–0.19 Gb$^3$). It is interesting to note that the magnitude of the activation energy $\Delta F_1$ for the annealed UFG steel decreases monotonously from 5.6 kT (0.70 Gb$^3$) at T = 800 °C up to 9.2 kT (1.16 Gb$^3$) at T = 900 °C (Fig. 19b). In our opinion, this result is related to the nucleation of the second phase particles in the course of heating up (Fig. 7).

So far, the formation of the long–range internal stress fields from the nonequilibrium grain boundaries, which prevent the free motion of the lattice dislocations (prevent the accommodative reconstruction of the defect structure) is the origin of the increased stress relaxation resistance of the UFG steel.

The change of the phase composition of the steel can be an additional factor increasing the stress relaxation resistance at ECAP. As follows from the analysis of the results of the XRD investigations, the stainless steel after N = 4 ECAP cycles contains from ~ 7–8% ($T_{ECAP}$ = 150 °C) up to 17–18% ($T_{ECAP}$ = 450 °C) of stronger α(δ)-phase particles. At the same external stress, the stress relaxation magnitude (the accommodation reconstruction of the defect structure) in the stronger α(δ)–phase will be smaller than the one in the γ-phase. In this connection, the increasing of the content of the stronger α(δ)–phase particles can promote the increase in the stress relaxation magnitude of the UFG steel.

Corrosion resistance

Analysis of the results of the corrosion tests demonstrated ECAP to result in an insufficient increase in the uniform corrosion rate $V_{corr}$ calculated according to the Tafel method. Besides, the analysis of the results of the electrochemical testing by the DLEPR method demonstrated the UFG steel to have a higher tendency to the IGC as compared to the coarse–grained steel. It should be stressed here that in spite of the increased tendency to the IGC, the UFG steel satisfies the requirements of GOST 9.914-91 in the resistance against IGC completely.

In our opinion, the increasing of the volume fraction of strain–induced martensite and, hence, the formation of the two–phase γ+α microstructure is the main origin of increased corrosion rate and of reduction of the resistance against to IGC in the UFG steels. The strain–induced martensite particles having a different chemical composition (unlike austenite) have a higher corrosion (dissolving) rate. Therefore, the increasing of the volume fraction of strain–induced martensite will lead to increasing of the uniform corrosion rate according to the ordinary rule: $V_{corr} = f_\gamma V_\gamma + f_\alpha V_\alpha$ where $V_\gamma$ and $V_\alpha$ are the dissolving rates for the $\gamma$– and $\alpha$–phases, respectively.

The formation of the two–phase microstructure leads to the appearing of the microgalvanic couples austenite – martensite in the material. These ones are the places of accelerated corrosion destruction during the electrochemical tests for IGC. So far, the increasing of the volume fraction of strain–induced martensite provides the conditions for the increase in the uniform corrosion rate and in the intergranular corrosion one.

The second factor promoting the reduction of the resistance of the stainless steel against IGC after ECAP can be the redistribution of the doping elements (chromium and nickel) during SPD. In [58], the grain boundaries in the nanocrystalline austenitic steel Fe-12%Cr-30%Ni with the grain size ~60 nm were shown to be enriched with nickel after SPD but to have a reduced chromium concentration. The width of the near-boundary zone enriched with nickel was predicted theoretically to increase with increasing temperature [58]. The strain-induced segregation of the Ni atoms at the austenite grain boundaries was utilized to explain the formation of the ferromagnetic clusters at the grain boundaries in the Fe-12%Cr-30%Ni and Fe-12%Cr-40%Ni steels during SPD [59]. Such a deformation-stimulated bundle of the solid solution would promote an accelerated electrochemical corrosion near the grain boundaries in the UFG steel Fe-18%Cr-10%Ni-0.1%Ti.

**Conclusions**

1. The samples of the UFG steel with improved mechanical properties were obtained by ECAP. After N = 4 ECAP cycles at 150 and 450 °C, the values of the ultimate strength of the steel were 1100 and 1020 MPa, respectively. The main contribution into the increasing of the strength of steel during ECAP is made by the increasing of the dislocation density and by the modification of the grain structure down to the submicron scale. The stages of the uniform strain flow were observed in the stress–strain tension curves σ(ε) of the specimens of the UFG steel at RT. The specimen fractures had a viscous character. The XRD phase analysis revealed the strain–induced martensite to form during ECAP. The strain–induced martensite content in the UFG steel microstructure achieves 17-18%.

2. The annealing of the UFG steel at the temperatures over 700 ºC leads to the beginning of the recrystallization processes, which is accompanied by the decreasing of the volume fraction of the strain–induced martensite and by the nucleation of the light–colored nanometer-sized particles, which presumably consist of σ–phase. The activation energy of the grain boundaries migration (6.0–8.3 $kT_m$) is 20-30% smaller than the one of the diffusion along the austenite grain boundaries. The reduction of the activation energy is caused by the presence of the excess density of defects – the orientation mismatch dislocations and of the products of the delocalization of these ones – at the nonequilibrium grain boundaries.

3. The UFG steel is featured by an improved stress relaxation resistance – by a higher macroelasticity stress and a smaller stress relaxation magnitude (at given magnitude of the stress applied). The increased stress relaxation resistance of the UFG steel is caused by a special internal stress relaxation mechanism related to the interaction of the lattice dislocations with the nonequilibrium grain boundaries in the UFG steel. The second probable origin of the increased stress relaxation resistance of the UFG steel can be the presence of stronger strain–induced martensite particles, which the accommodation reconstruction of the dislocation structure is difficult in.

4. The ECAP process leads to the reduction of the corrosion resistance of the austenite steel – the increase in the uniform corrosion rate and the increasing of the tendency of the steel to the intergranular corrosion were observed. The reduction of the corrosion resistance is caused, first of all, by the presence of the strain–induced martensite particles, which have a greater dissolving rate. The presence of the strain–induced martensite particles leads to the appearance of the microgalvanic couples martensite – austenite in the steel microstructure, at the grain boundaries of which an accelerated intergranular corrosion is possible.

### Acknowledgements

The present study was supported by Ministry of Science and Higher Education of the Russian Federation (Grant No. 0729-2020-0060).

The TEM microstructure were carried out the equipment of shared research facility "Materials Science and Metallurgy" NUST "MISIS" funded by Ministry of Science and Higher Education of the Russian Federation (Project No. 075-15-2021-696).

### Author Contribution Statement

V.N. Chuvil'deev - Project administration, Writing - review & editing, Funding acquisition

A.V. Nokhrin – Investigation (fractography, SEM) & Analysis of experimental results & Writing of manuscript & Data curation

N.A. Kozlova – Investigation (corrosion test)

M.K. Chegurov – Investigation (corrosion test, fractography)

M.Yu. Gryaznov, S.V. Shotin – Investigation (mechanical tensile tests)

V.I. Kopylov – Investigation (obtained by UFG steel by ECAP, optimization of ECAP modes)

N.V. Melekhin – Investigation (relaxation test)

C.V. Likhnitskii – Investigation (microhardness, optical microscopy)

N.Yu. Tabachkova - Investigation (TEM)

**Conflict of interest**. The authors declare that they have no conflict of interest.

Table 1. Results of mechanical testing. The magnitudes of the ultimate strength ($\sigma_b$, MPa) and of the elongation to failure ($\delta$, %) for the samples of stainless steel for the tension tests at different temperatures

| $T_{test}$, °C | CG steel | | N = 2 | | N = 3 | | | | N = 4 | | | |
|---|---|---|---|---|---|---|---|---|---|---|---|---|
| | | | $T_{ECAP}$=450 °C | | $T_{ECAP}$=150 °C | | $T_{ECAP}$=450 °C | | $T_{ECAP}$=150 °C | | $T_{ECAP}$=450 °C | |
| | $\sigma_b$ | $\delta$ | $\sigma_b$ | $\delta$ | $\sigma_b$ | $\delta$ | $\sigma_b$ | $\delta$ | $\sigma_b$ | $\delta$ | $\sigma_b$ | $\delta$ |
| RT | 720 | 125 | 950 | 70 | 1100 | 40 | 950 | 65 | 1100 | 45 | 1020 | 60 |
| 450 | 420 | 65 | 800 | 40 | 870 | 35 | 720 | 20 | 920 | 22 | 760 | 30 |
| 600 | 350 | 65 | 650 | 50 | 600 | 50 | 600 | 48 | 630 | 45 | 640 | 45 |
| 750 | 250 | 70 | - | | 120 | 250 | 290 | 105 | 240 | 185 | 290 | 120 |
| 800 | 220 | 75 | 250 | 110 | 150 | 200 | 200 | 150 | 152 | 220 | 205 | 160 |
| 900 | - | - | - | - | - | - | 98 | 190 | - | - | - | - |

Table 2. Microhardness of the samples of steel after tension testing at different temperatures. The mean recrystallized grain sizes [d, μm] for the samples tested at 800 and 900 °C are given in braces

| $T_{test}$, °C | Microhardness ($H_v$, GPa) | | | | | | | | | | | |
|---|---|---|---|---|---|---|---|---|---|---|---|---|
| | CG steel | | N = 2 | | N = 3 | | | | N = 4 | | | |
| | | | $T_{ECAP}$=450°C | | $T_{ECAP}$=150°C | | $T_{ECAP}$=450°C | | $T_{ECAP}$=150°C | | $T_{ECAP}$=450°C | |
| | Zone I | Zone II | Zone I | Zone II | Zone I | Zone II | Zone I | Zone II | Zone I | Zone II | Zone I | Zone II |
| RT | 2.15 | 3.71 | 3.49 | 4.34 | 3.91 | 4.45 | 3.47 | 4.38 | 3.99 | 4.50 | 3.51 | 4.46 |
| 450 | 1.87 | 2.95 | - | - | - | - | 3.64 | 3.55 | 4.07 | 4.13 | 3.76 | 3.75 |
| 600 | 1.88 | 2.68 | 2.94 | 3.38 | 4.05 | 3.63 | 3.56 | 3.49 | 4.25 | 3.75 | 3.77 | 3.60 |
| 750 | 1.76 | 2.38 | - | - | - | - | 2.79 | 2.77 | 3.33 | 2.52 | 3.38 | 2.71 |
| 800 | 1.84 (33) | 2.26 (41) | 2.40 (21.5) | 2.54 (2.0) | 2.13 (2.7) | 2.39 (2.3) | 2.15 (2.9) | 2.48 (1.6) | 2.96 (2.5) | 3.05 (1.9) | 2.21 (2.6) | 2.29 (1.6) |
| 900 | - | - | - | - | - | - | 1.90 (6.8) | 1.99 (4.4) | - | - | - | - |

Note: Zone I is the non-deformed one, Zone II is the deformed one (the destruction zone)

Table 3. Results of the electrochemical corrosions testing of the coarse-grained and UFG steel

| Steel | Tafel test results | | | DLEPR test results (GOST 9.914-91) | | IGC test (GOST 6232-2003) |
|---|---|---|---|---|---|---|
| | $E_{corr}$, mV | $i_{corr}$, mA/cm$^2$ | $V_{corr}$, mm/year | $S_1/S_2$, $10^4$ | Corrosion character | Corrosion character |
| Initial state | -403 | 0.073 | 0.58 | 0.93 | IGC | IGC or PC |
| ECAP, N = 1, T = 150°C | -402 | 0.072 | 0.56 | 1.64 | UC | PC |
| ECAP, N = 2, T = 150°C | -403 | 0.083 | 0.64 | 1.96 | UC | PC |
| ECAP, N = 3, T = 150°C | -404 | 0.084 | 0.65 | 2.07 | UC | - |
| ECAP, N = 4, T = 150°C | -404 | 0.084 | 0.65 | 2.34 | UC | - |
| ECAP, N = 1, T = 450°C | -404 | 0.092 | 0.71 | 2.78 | UC | PC |
| ECAP, N = 2, T = 450°C | -406 | 0.084 | 0.64 | 3.25 | UC | - |
| ECAP, N = 3, T = 450°C | -406 | 0.099 | 0.77 | 2.41 | UC | - |
| ECAP, N = 4, T = 450°C | -403 | 0.097 | 0.75 | 2.22 | UC | - |

Note: IGC – intergranular corrosion, UC – uniform corrosion, PC – pitting corrosion

**List of Figures**

Figure 1 – Microstructure of stainless steel in the initial state: (a, b, c, d); the δ-phase nuclei in steel in the initial state (a, b – optical microscopy; c, d – SEM); (e, f) microstructure of the austenite grains. TEM

Figure 2. Macrostructure of the steel specimens after the first ECAP cycle at 150 (a) and 450 °C (b)

Figure 3 – Results of XRD phase analysis of the steel samples in the initial state and after ECAP: (a) XRD curves for the steel samples after different numbers of ECAP cycles at T = 450 °C; (b) dependence of the mass fraction of the α–phase on the number of the ECAP cycles at 150 and 450 °C

Figure 4. Microphotographs (a, c, d, e) and the electron diffraction patterns (b, f) of the steel microstructure after ECAP (N = 4) at 150 °C (a, b, c, d) and 450 °C (e, f)

Figure 5. Dependencies of the grain sizes on the annealing temperature for the UFG steel specimens subjected to ECAP at $T_{ECAP}$ = 450 °C

Figure 6. Nucleation of the second phase particles at the heating up of the UFG steel samples (N = 4, $T_{ECAP}$ = 450 °C): a – initial state; b – heating up to 500 °C, holding 60 min; c – heating up to 600 °C, holding 60 min; d – heating up to 700 °C, holding 60 min; e – heating up to 800 °C, holding 30 min; f – heating up to 800 °C, holding 60 min

Figure 7. Enlarged image of the nucleation of the second phase particles in the UFG steel (N = 4, $T_{ECAP}$ = 450 °C) after at heating up to 800 °C and holding for 60 min. The regions of intensive particle nucleation are marked by the dashed lines

Figure 8: Results of investigations of the mechanical properties of UFG steel ($T_{ECAP}$ = 450 °C): a) dependencies of the mean grain sizes and of the mechanical properties of steel on the number of ECAP cycles; b) dependence of the yield strength on the grain size in the axes $\sigma_y - d^{-1/2}$ axes

Figure 9. Dependencies of the macroelasticity stress (a) and of the yield strength (b) on the temperature of the 1-hour annealing of UFG steel

Figure 10 – Stress–strain tension curves for the CG and UFG steel samples at RT

Figure 11. Fractographic analysis of the fractures of steel after the tension tests at RT: (a, b) coarse grained steel, (c, d) UFG steel (N = 4, $T_{ECAP}$ = 150 °C), (e, f) UFG steel (N = 4, $T_{ECAP}$ = 450 °C). In Figs. 11a, b: Zone 1 – the fibrous fracture zone; Zone 2 – the radial zone; Zone 3 – the cut zone; in Fig. 11d – the fibrous zone consisting of a set of pits and featuring the viscous destruction

Figure 12 Results of the stress relaxation tests: (a) the stress relaxation curves of the CG and UFG steel specimens; (b) the stress relaxation curves of the UFG steel specimens (ECAP, N = 1, 150 °C) after annealing at different temperatures

Figure 13. Tension diagram for the CG (a) and UFG (b, c) steel at elevated temperatures: (a) CG steel; (b) UFG steel (N = 4, 150 °C), (c) UFG steel (N = 4, 450 °C)

Figure 14. Fractographic analysis of the steel fractures after the superplasticity tests at 600 °C: (a, b) CG steel, (c, d) UFG steel (N = 4, $T_{ECAP}$ = 150 °C), (e, f) UFG steel (N = 4, $T_{ECAP}$ = 450 °C). Designations in the figures are the same as in Figure 11

Figure 15. Results of the microstructure investigations of the nondeformed areas (a, c, e) and of the deformed ones (b, d, f) of the specimens after the tension tests at 800 °C: (a, b) CG steel; (c, d) UFG steel (ECAP, N = 3, 150 °C); (e, f) UFG steel (ECAP, N = 4, 450 °C)

Figure 16. Results of the electrochemical investigations of the coarse–grained and UFG steel samples: a) Tafel curves ln(i)–E; b) results of the DLEPR tests

Figure 17. Surfaces of the CG steel specimen (a) and of the UFG one (N = 4, 450 °C) (b) after the DLEPR tests according to GOST 9.914-91

Figure 18. The surfaces of the CG steel specimen (a) and of the UFG one (N = 4, 450 °C) after testing in boiling acid solution according to GOST 6232-2003

Figure 19. Dependencies of the stress relaxation magnitude on the stress applied in the $\ln(\Delta\sigma) - 1-\sigma/\sigma^*$ axes: (a) comparison of the course grained and UFG steels (analysis of the data presented in Fig. 12a); (б) effect of the annealing temperature on the relaxation curves for UFG steel (analysis of the data presented in Fig. 12b)

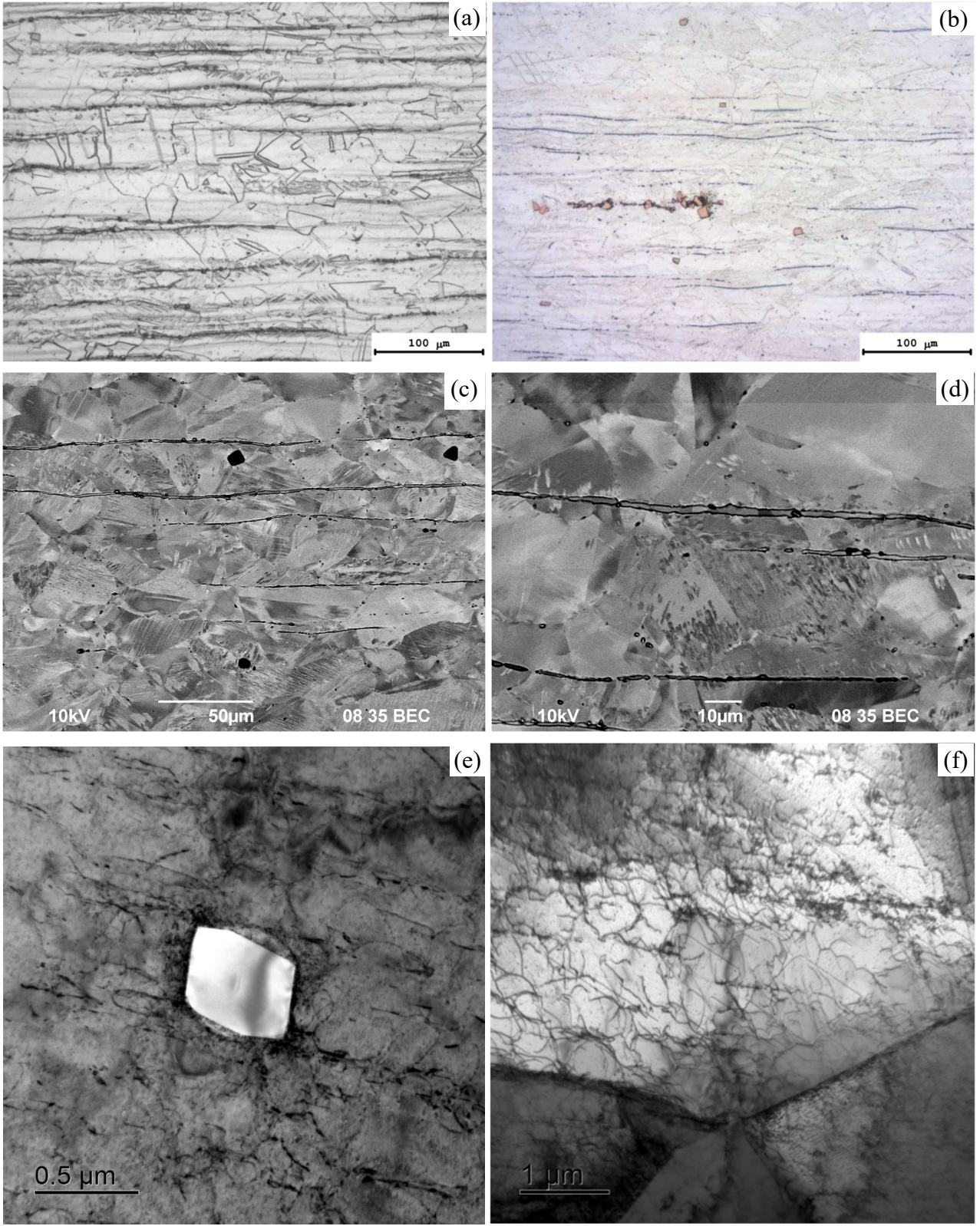

Figure 1

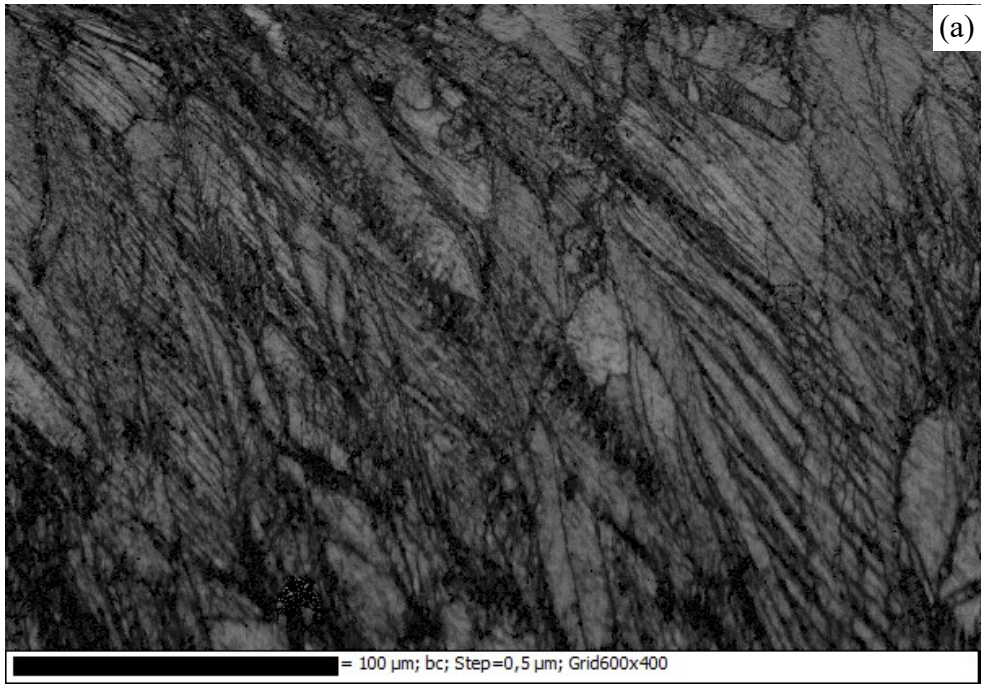
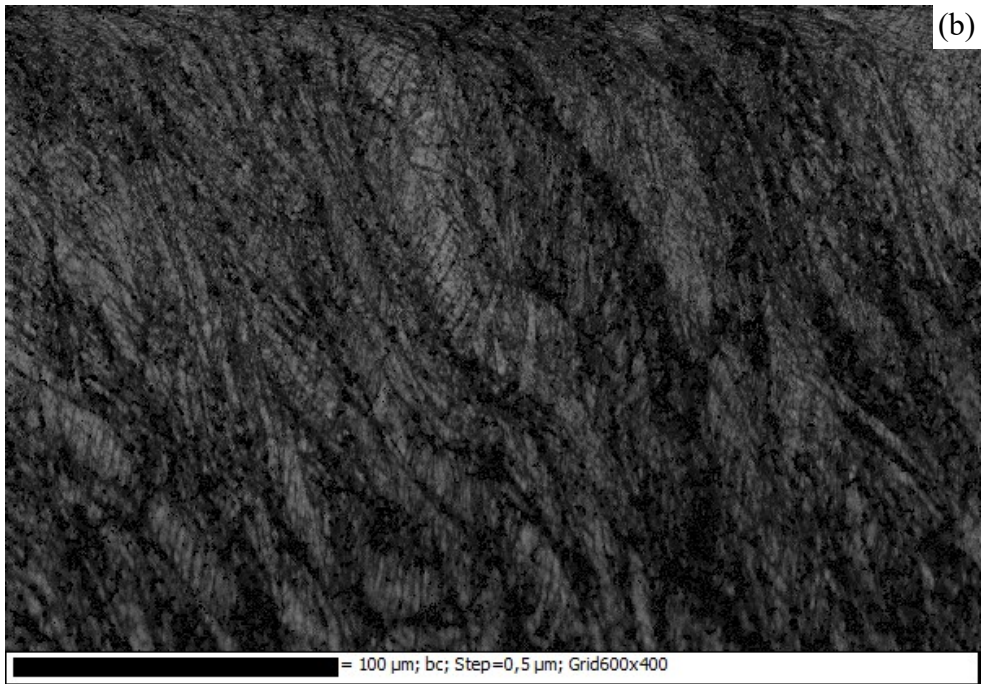

Figure 2

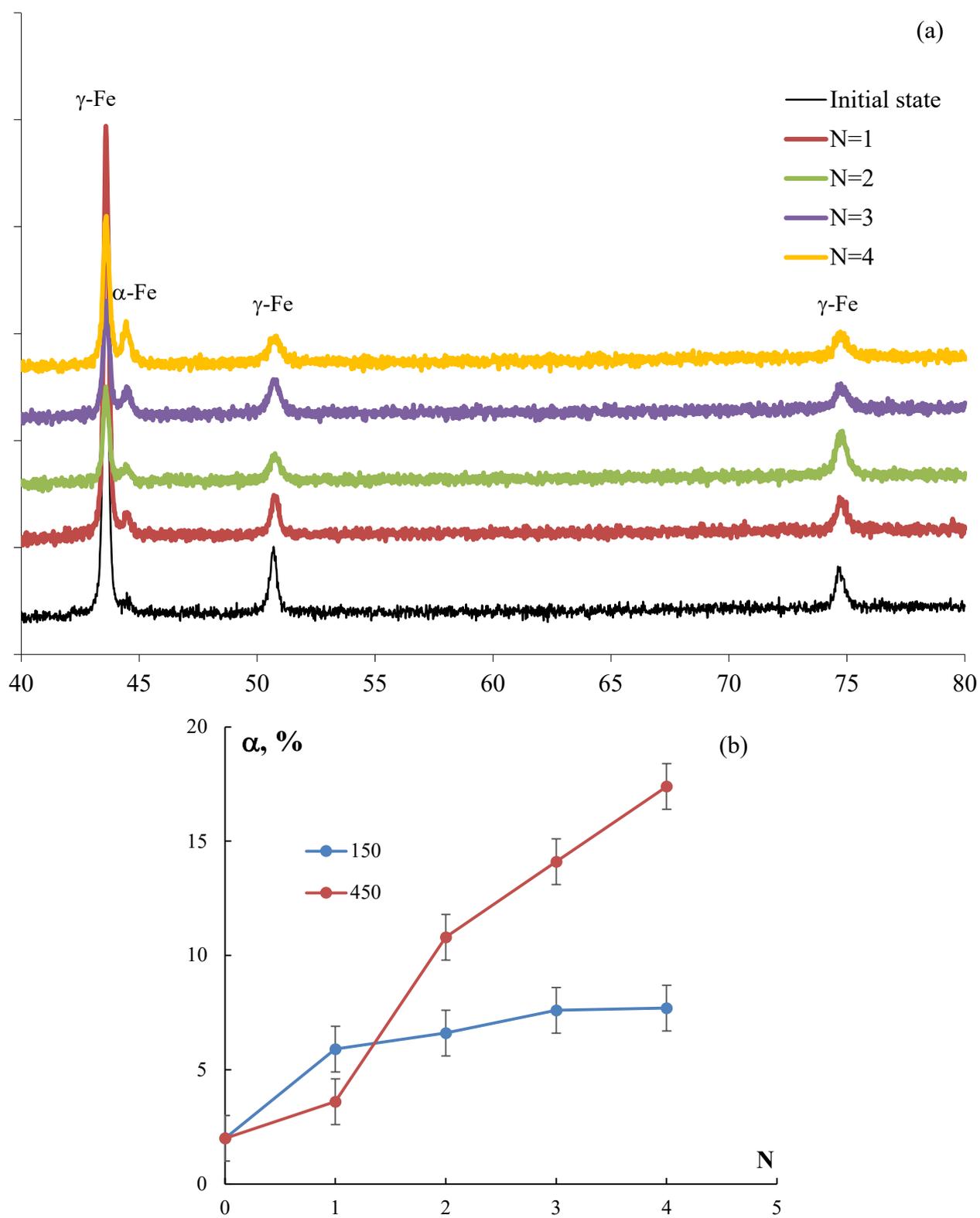

Figure 3

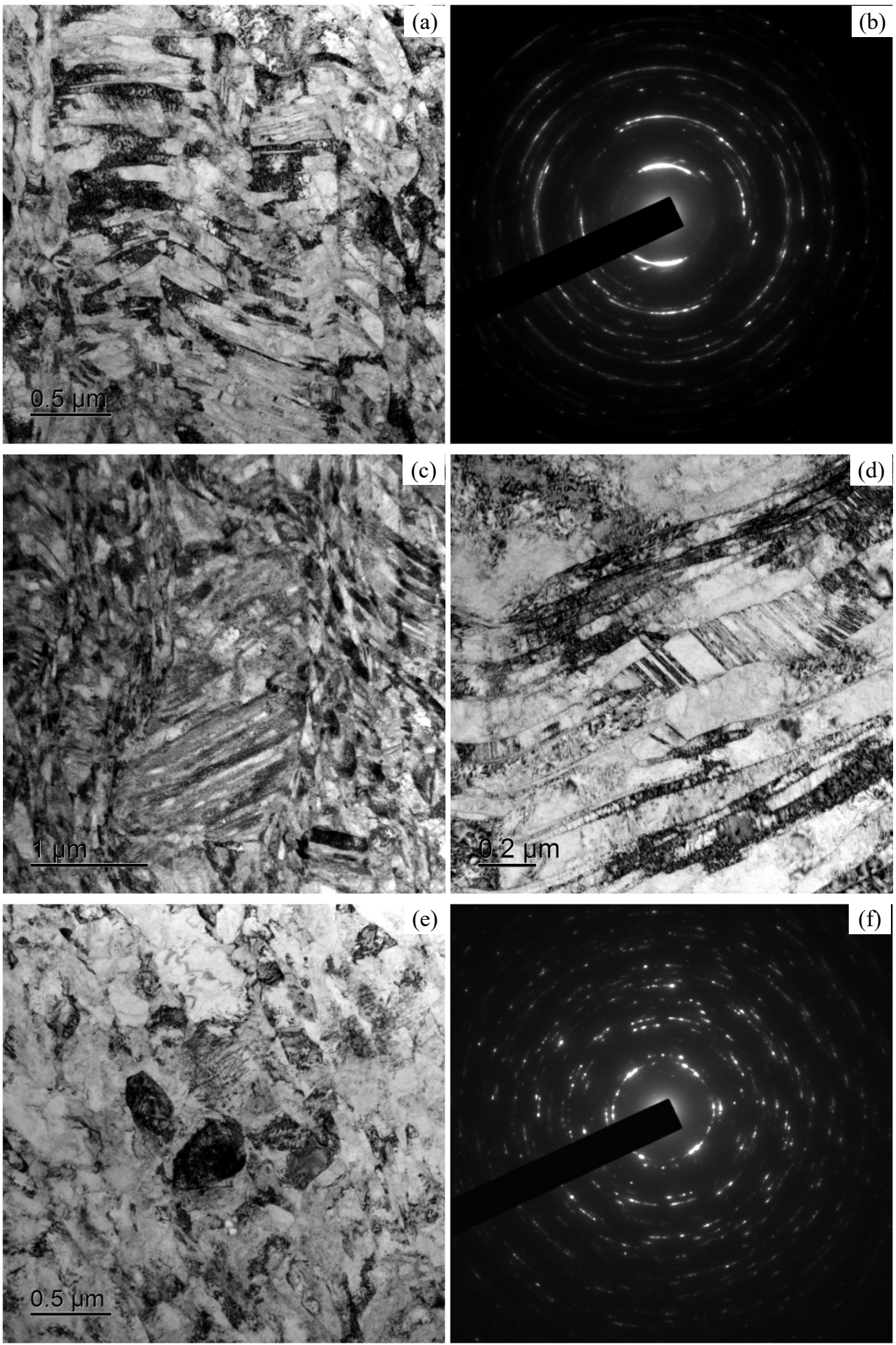

Figure 4

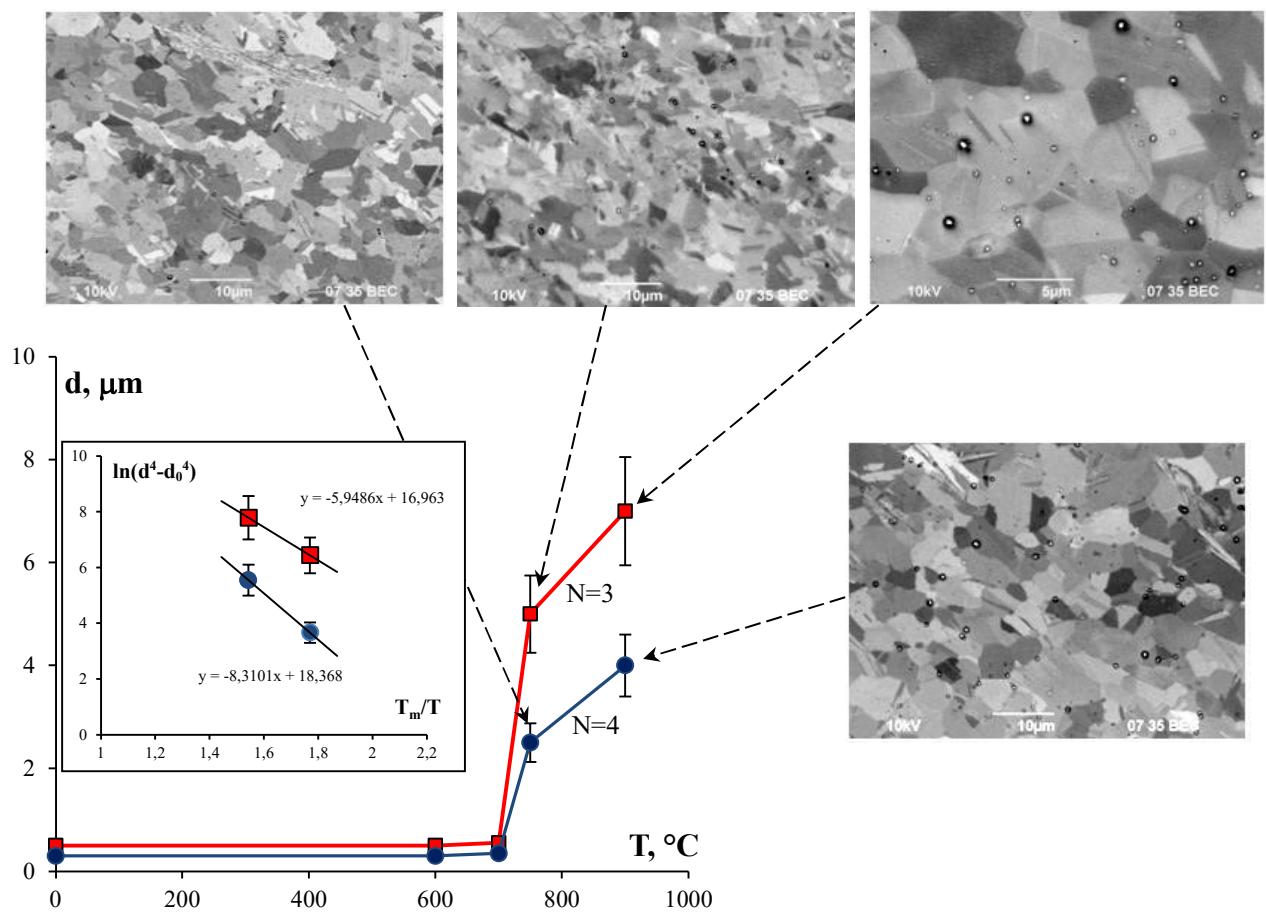

Figure 5

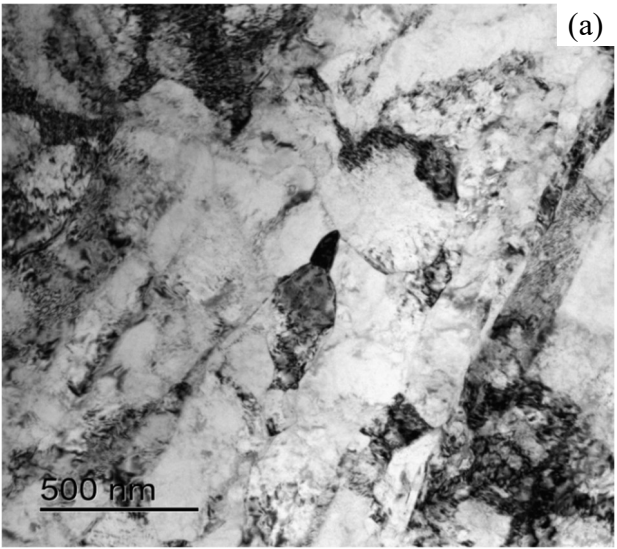
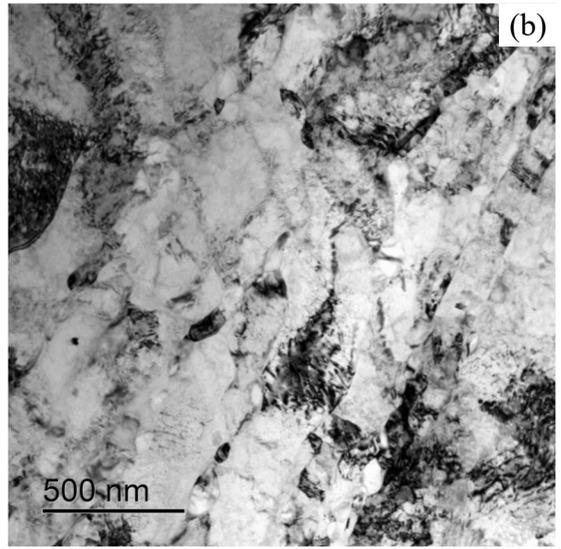
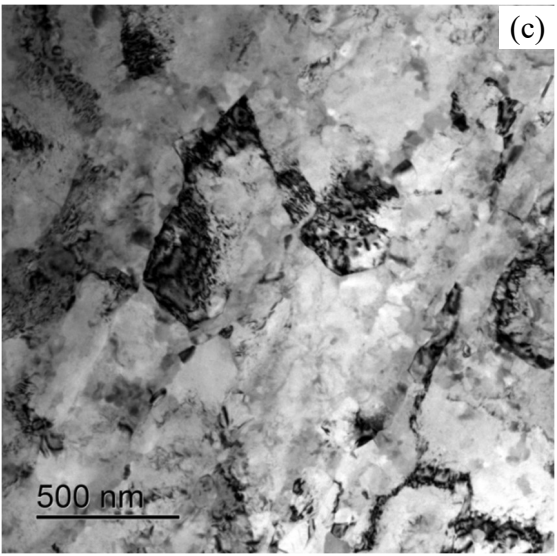
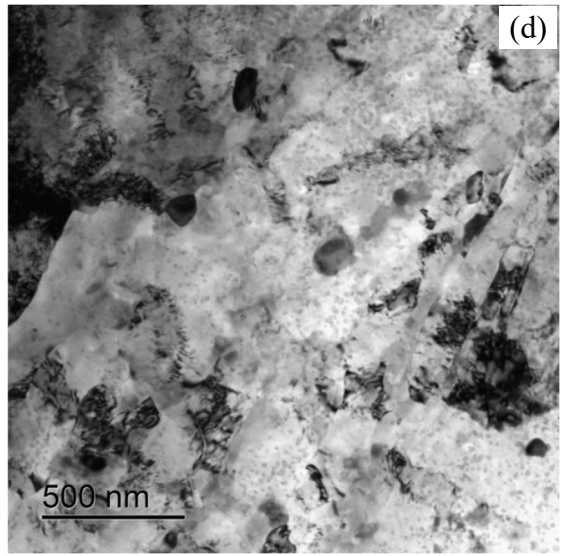
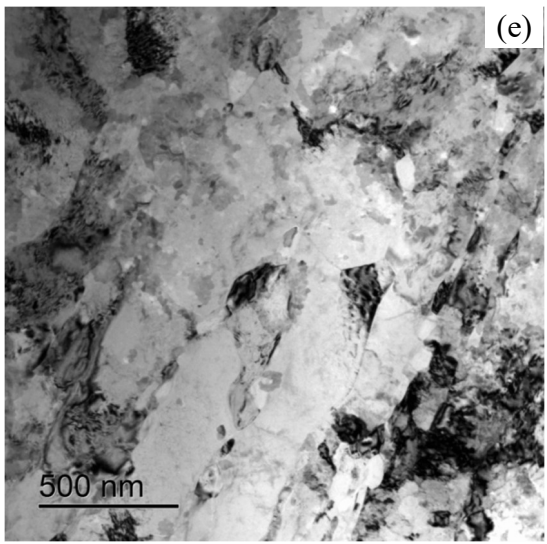
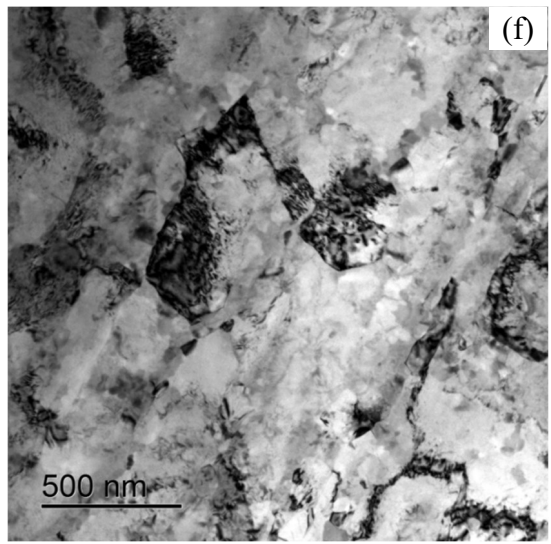

Figure 6

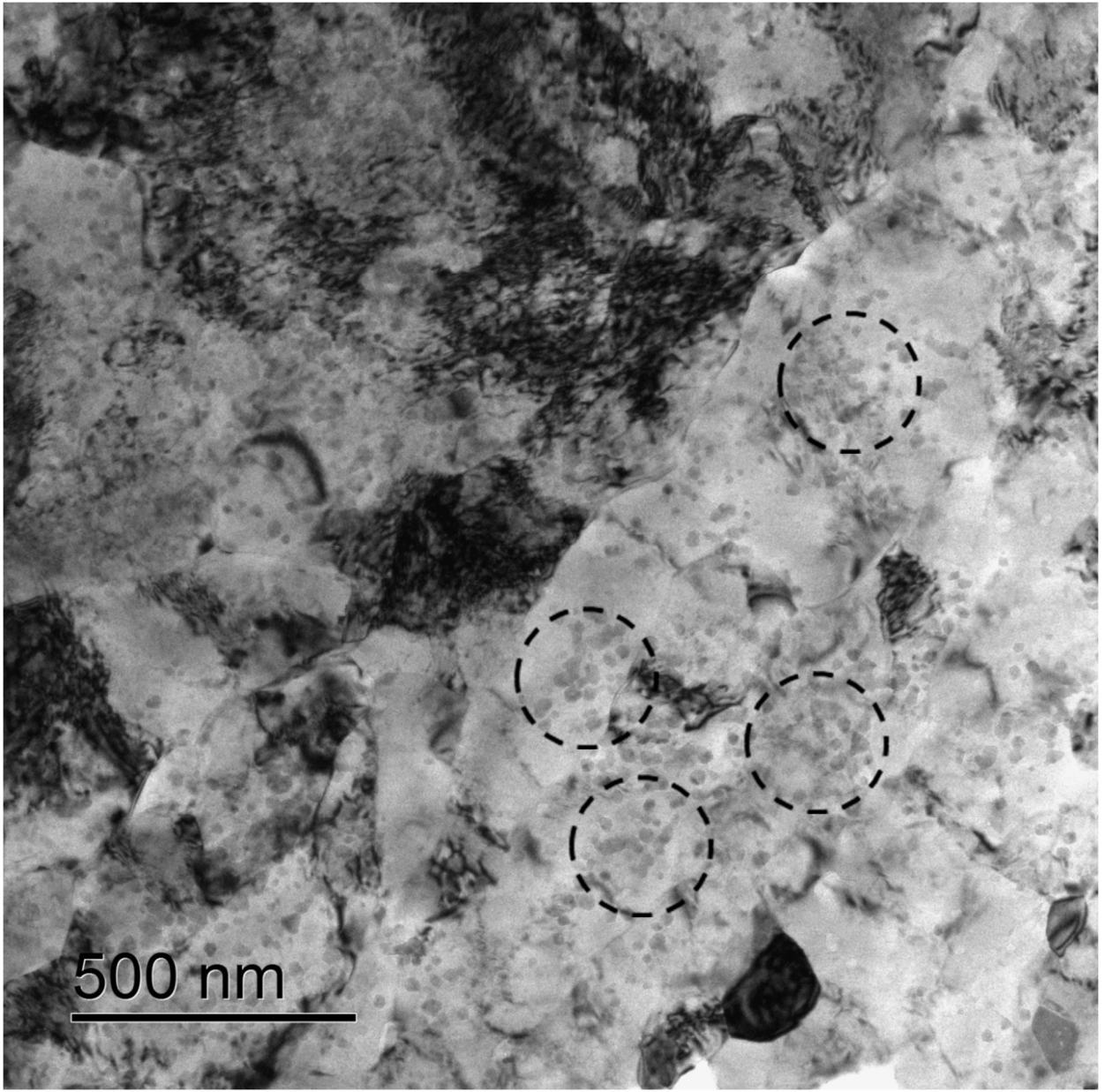

Figure 7

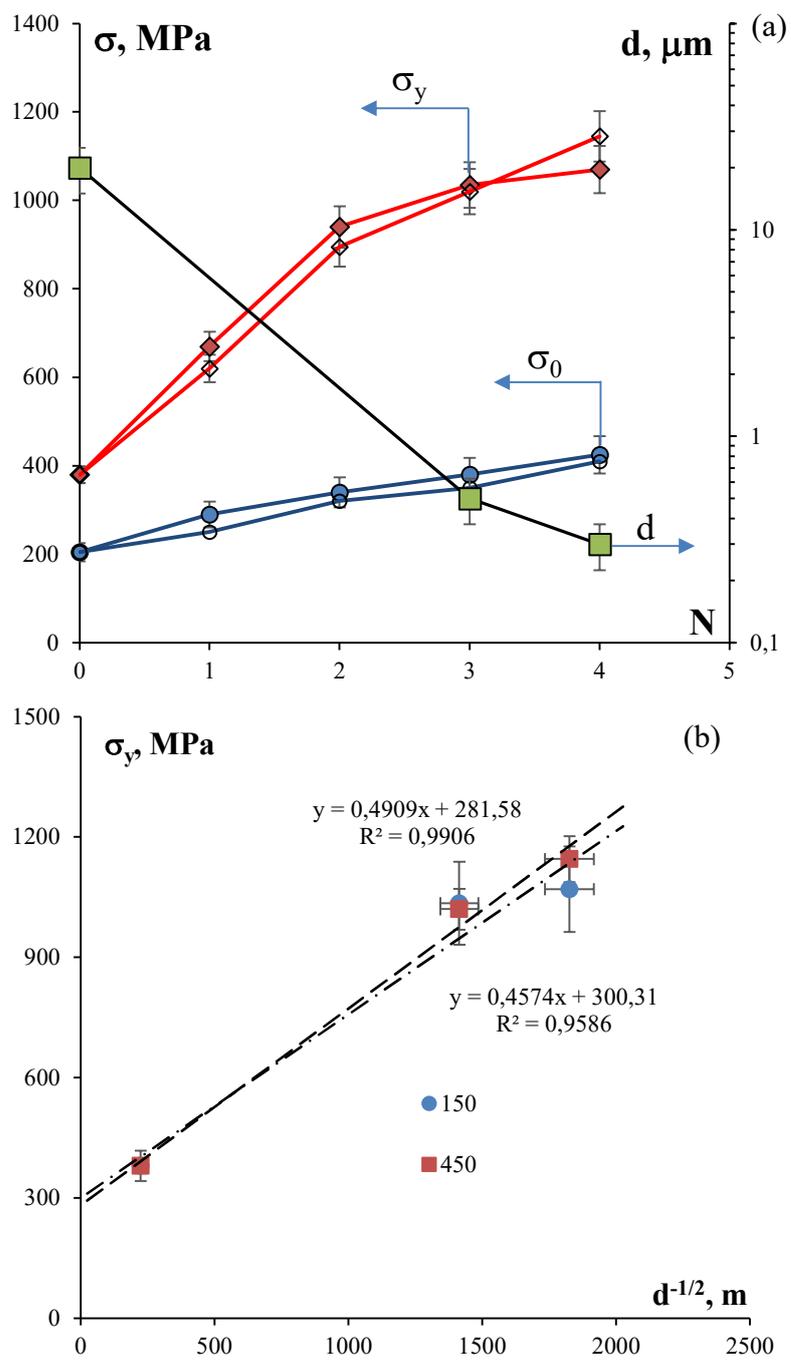

Figure 8

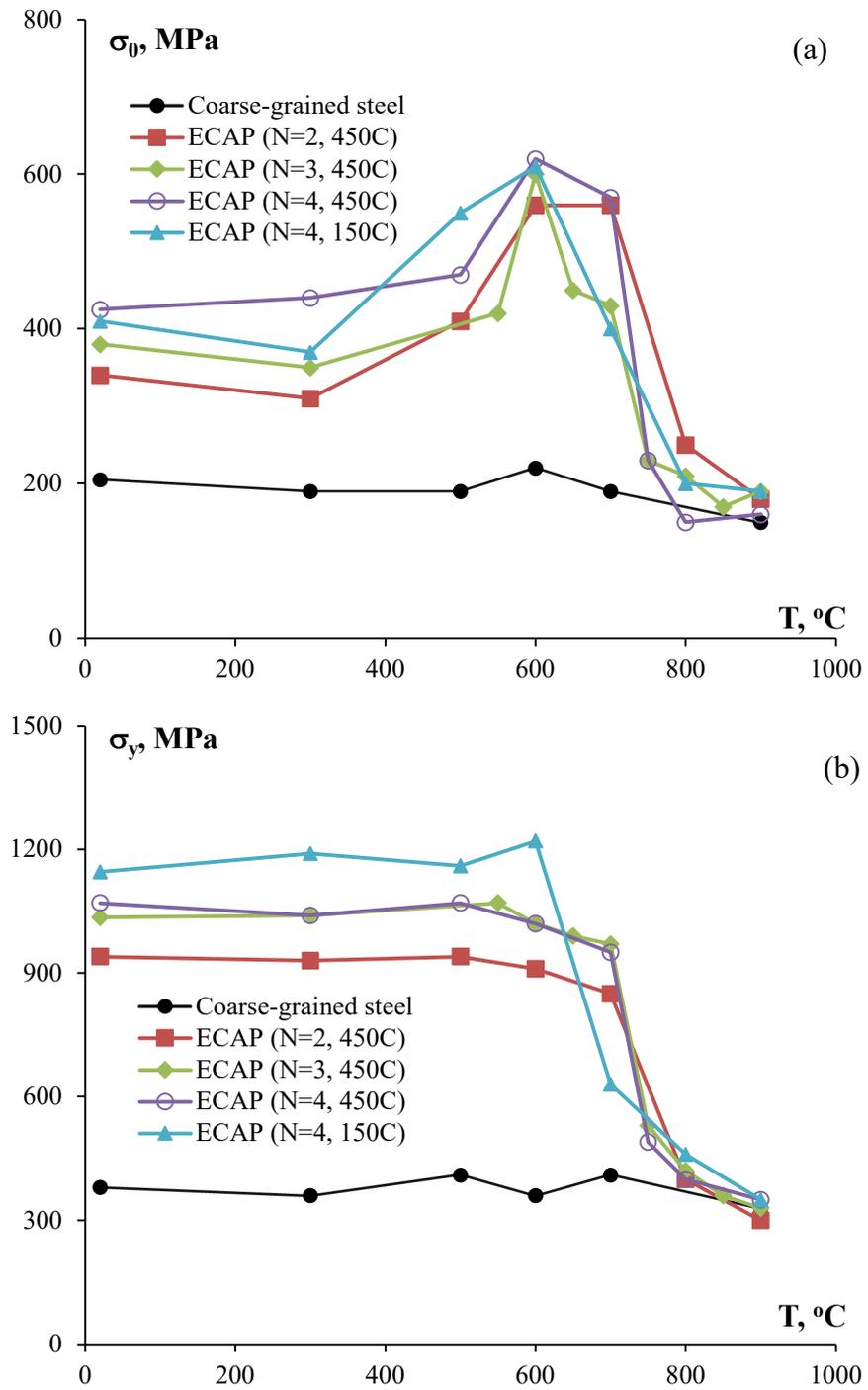

Figure 9

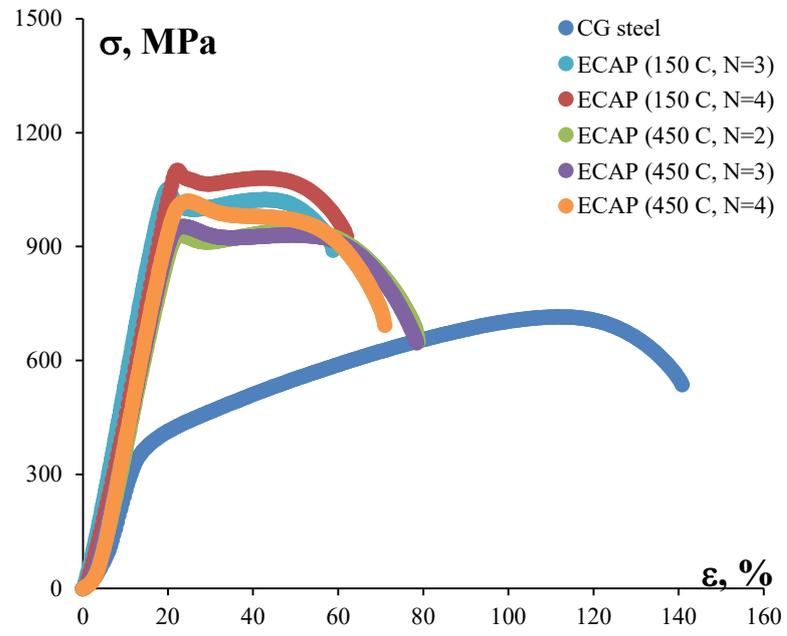

Figure 10

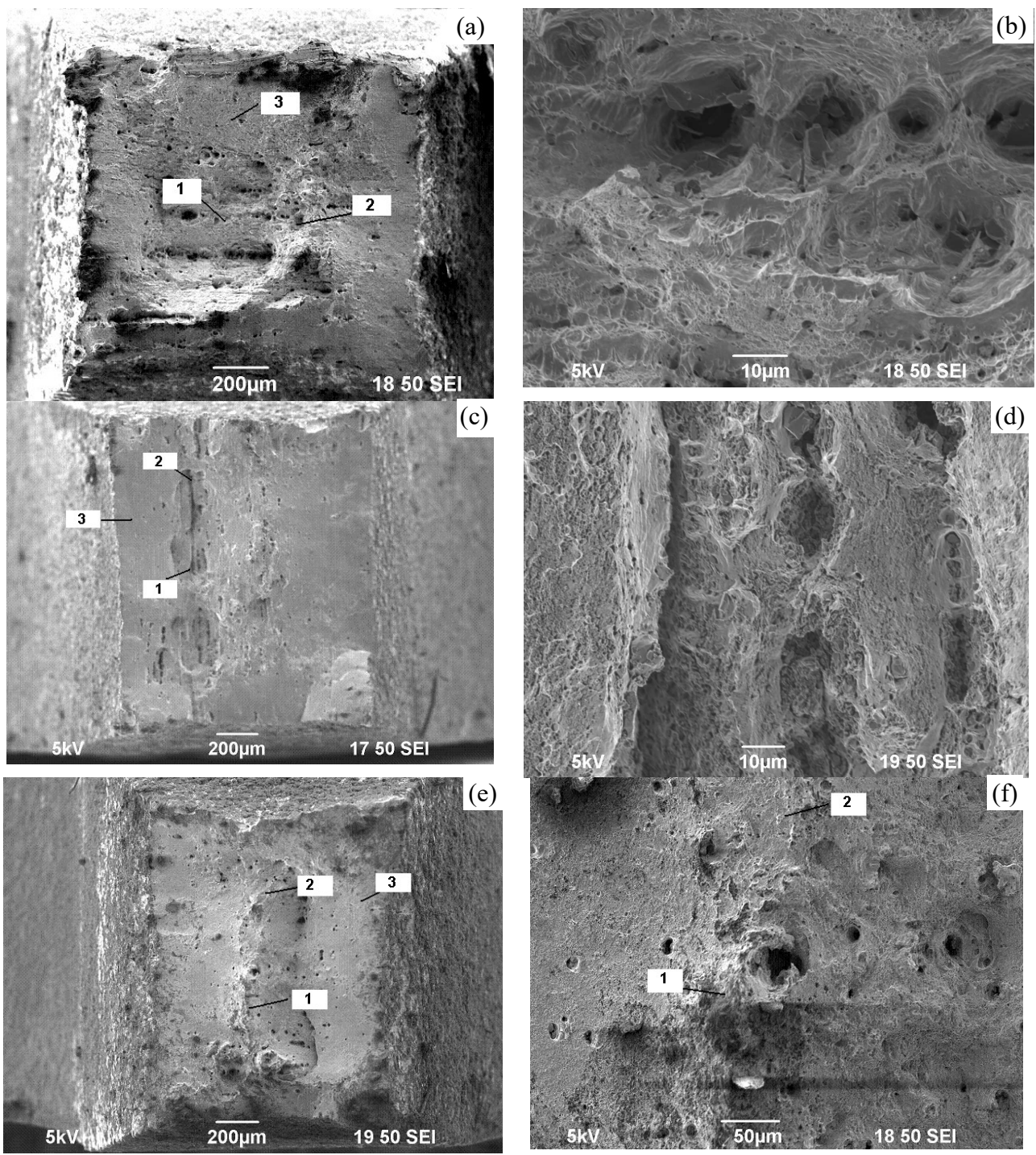

Figure 11

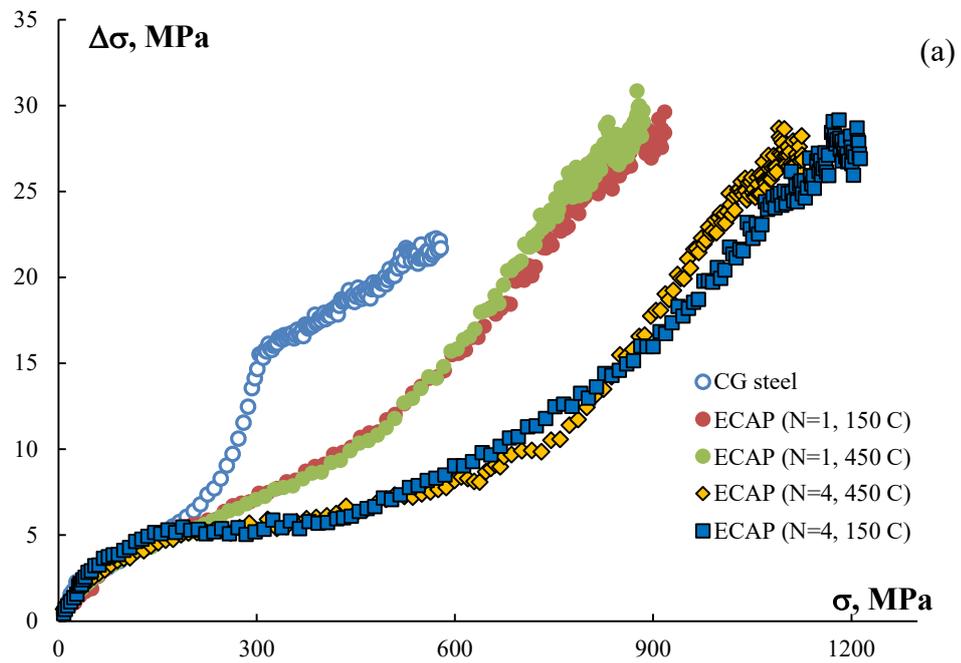

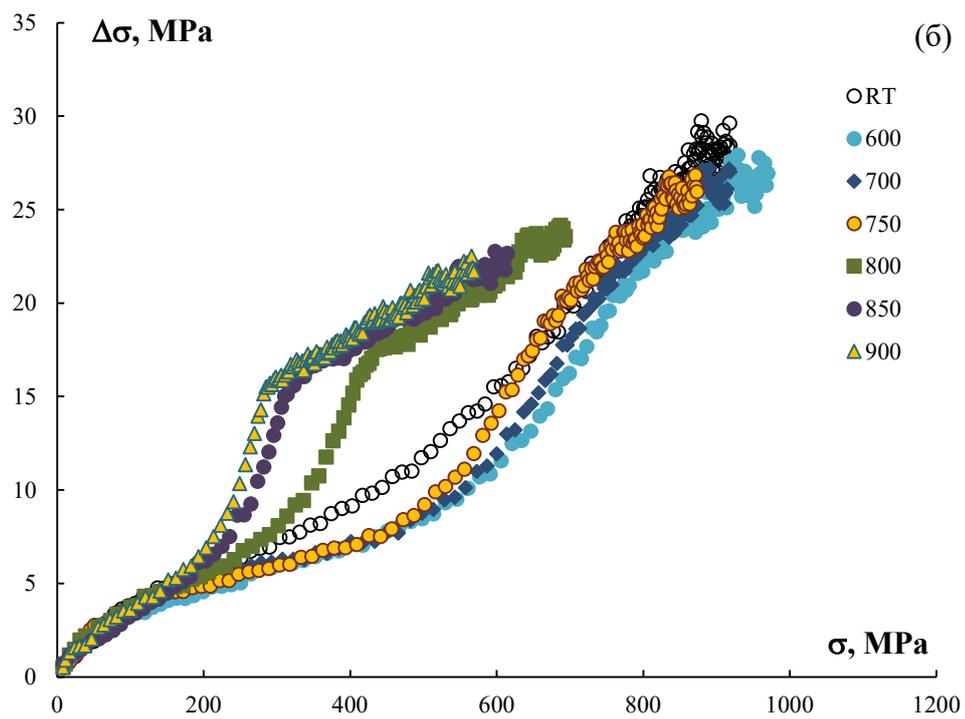

Figure 12

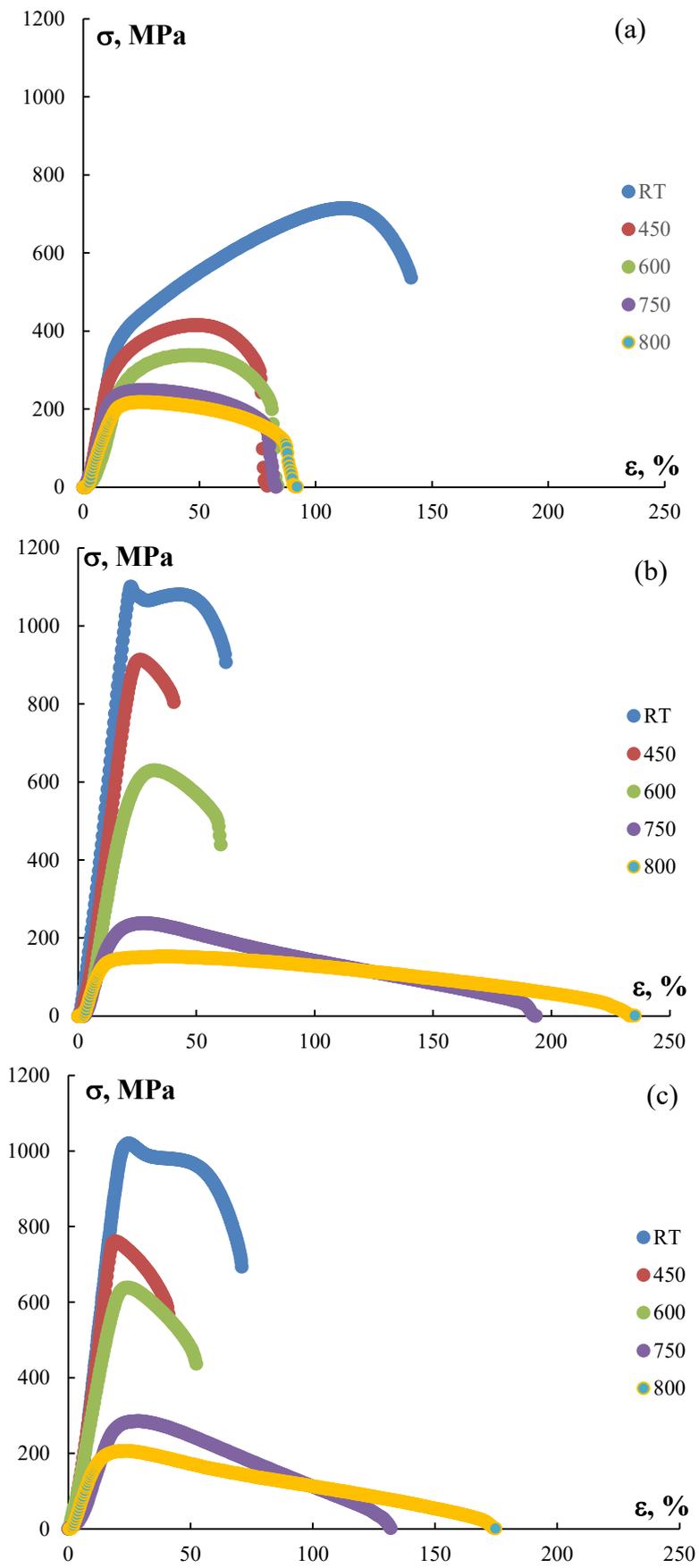

Figure 13

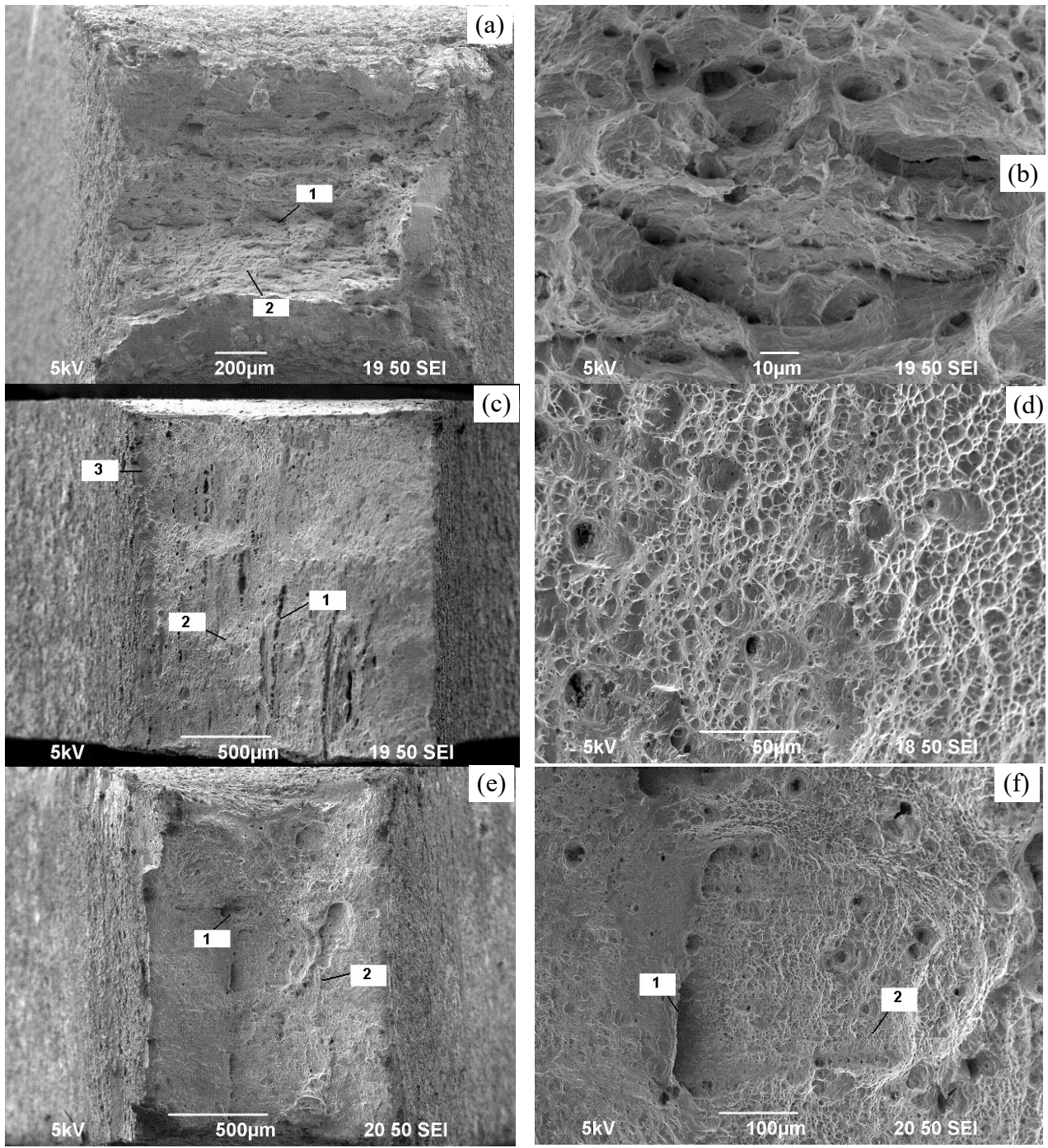

Figure 14

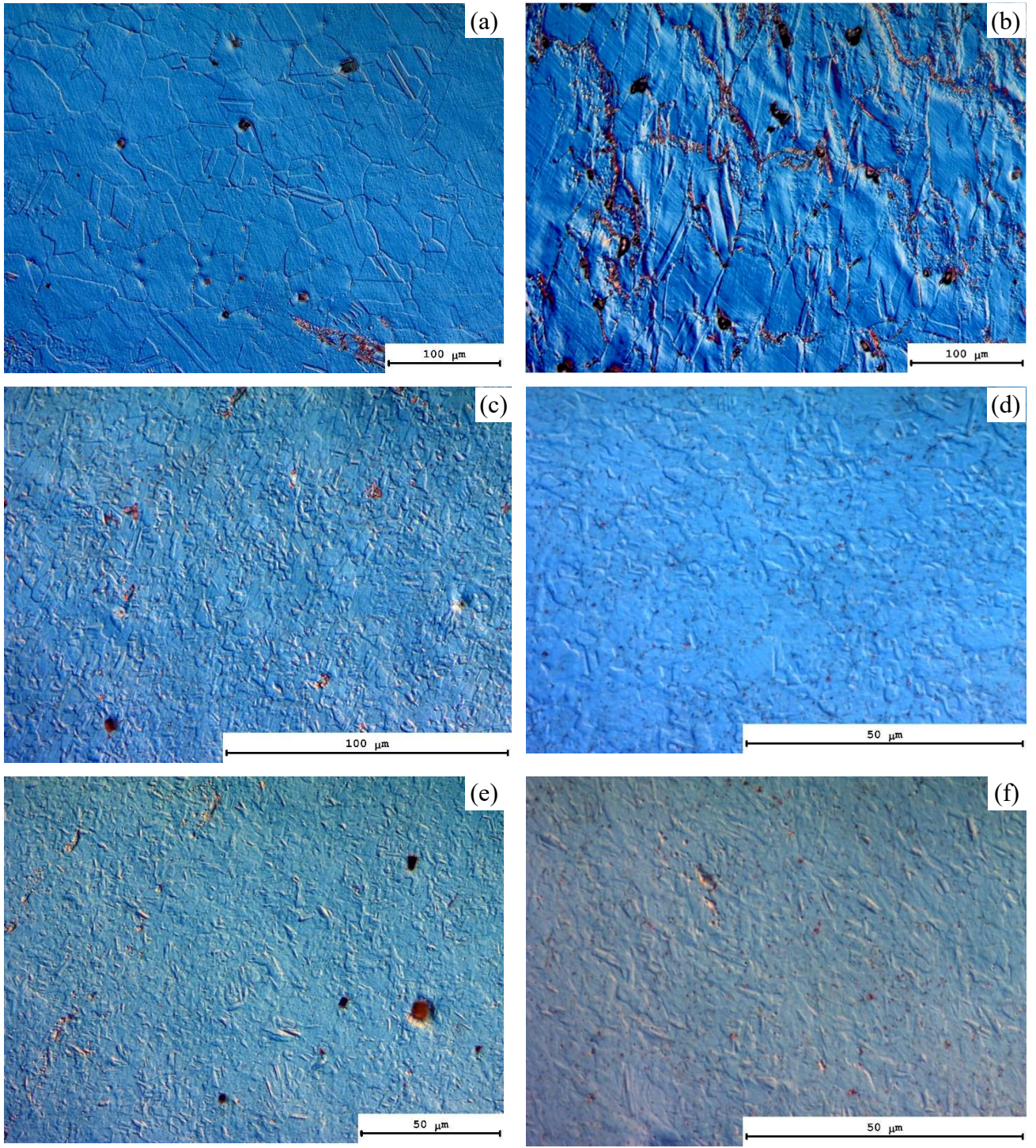

Figure 15

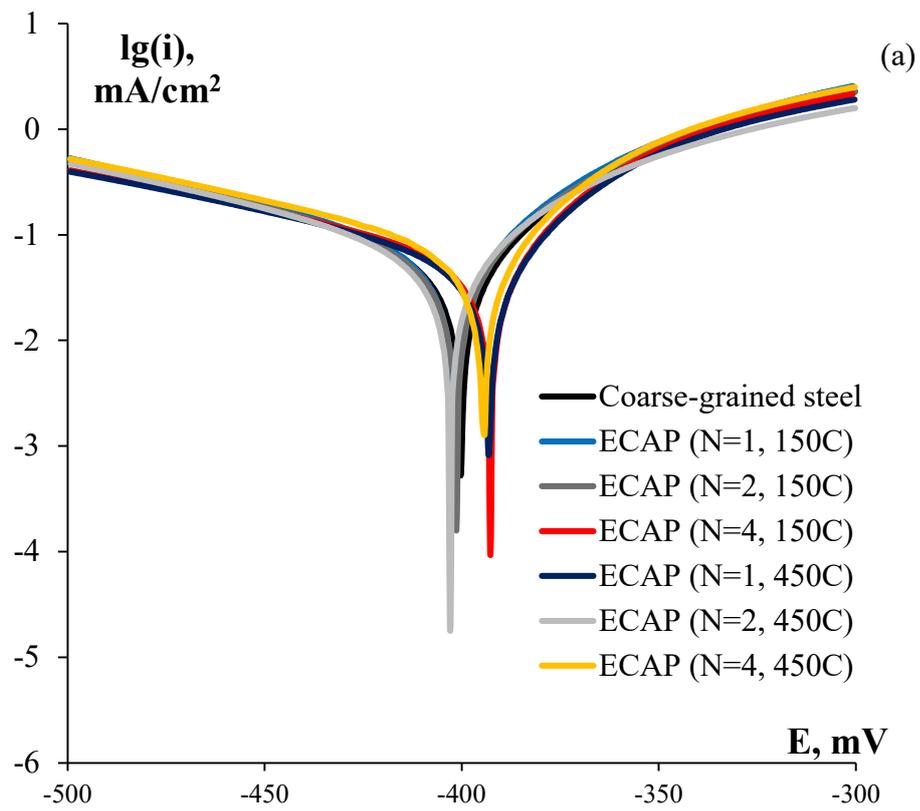

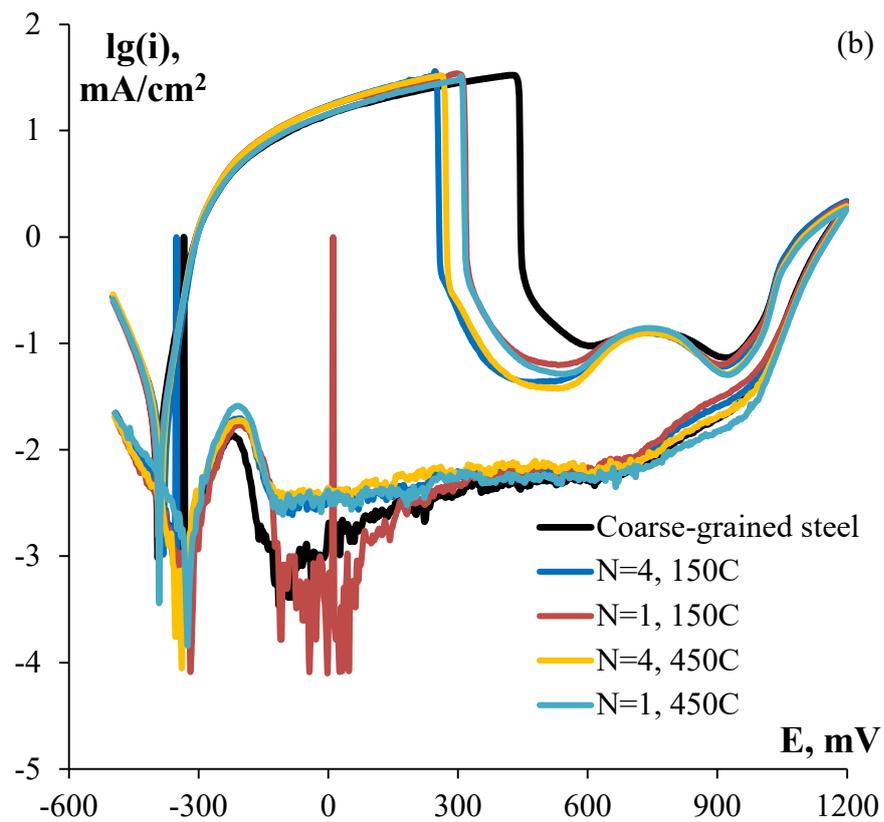

Figure 16

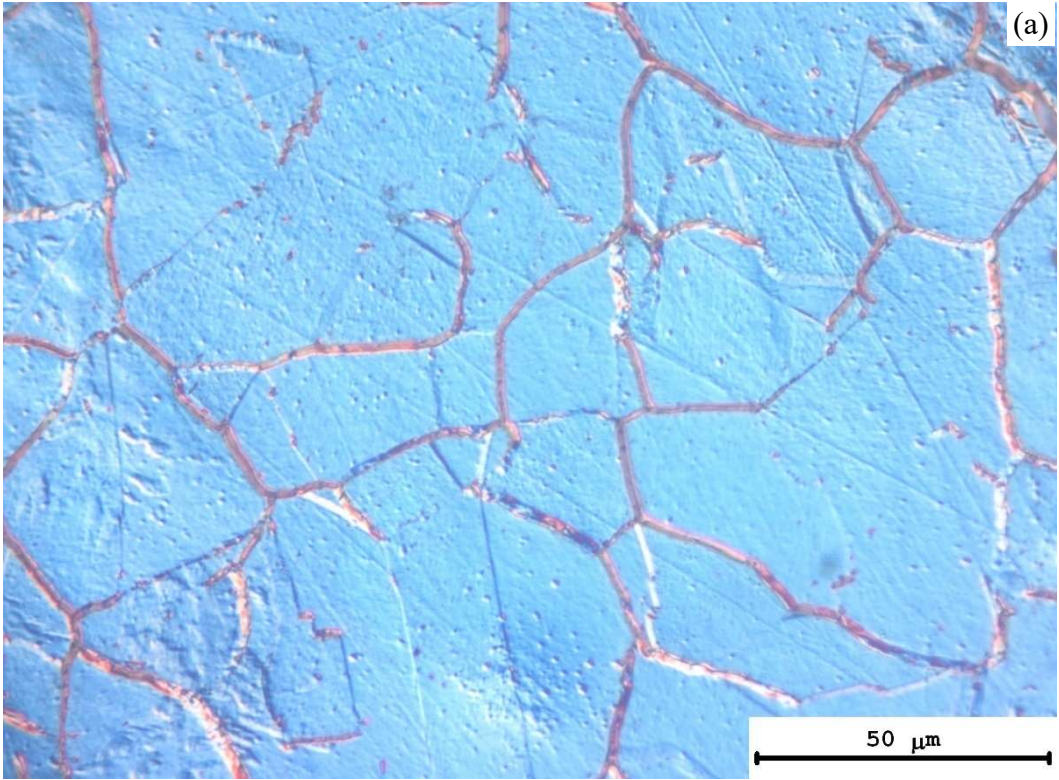
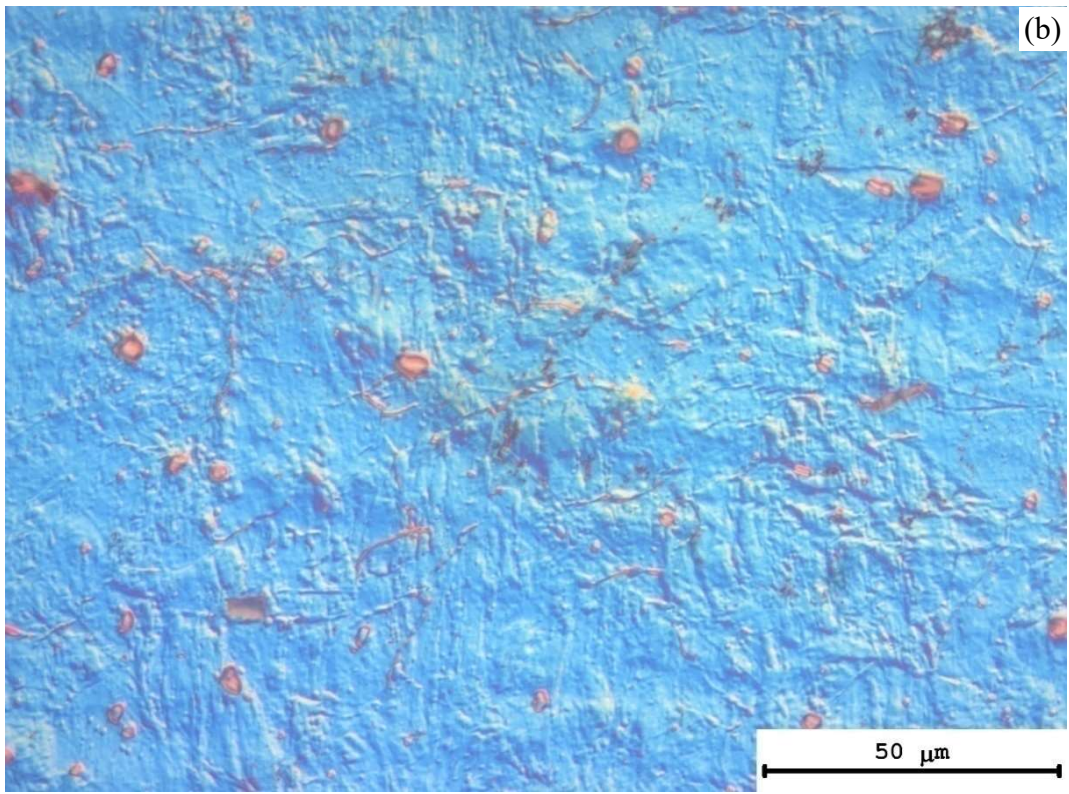

Figure 17

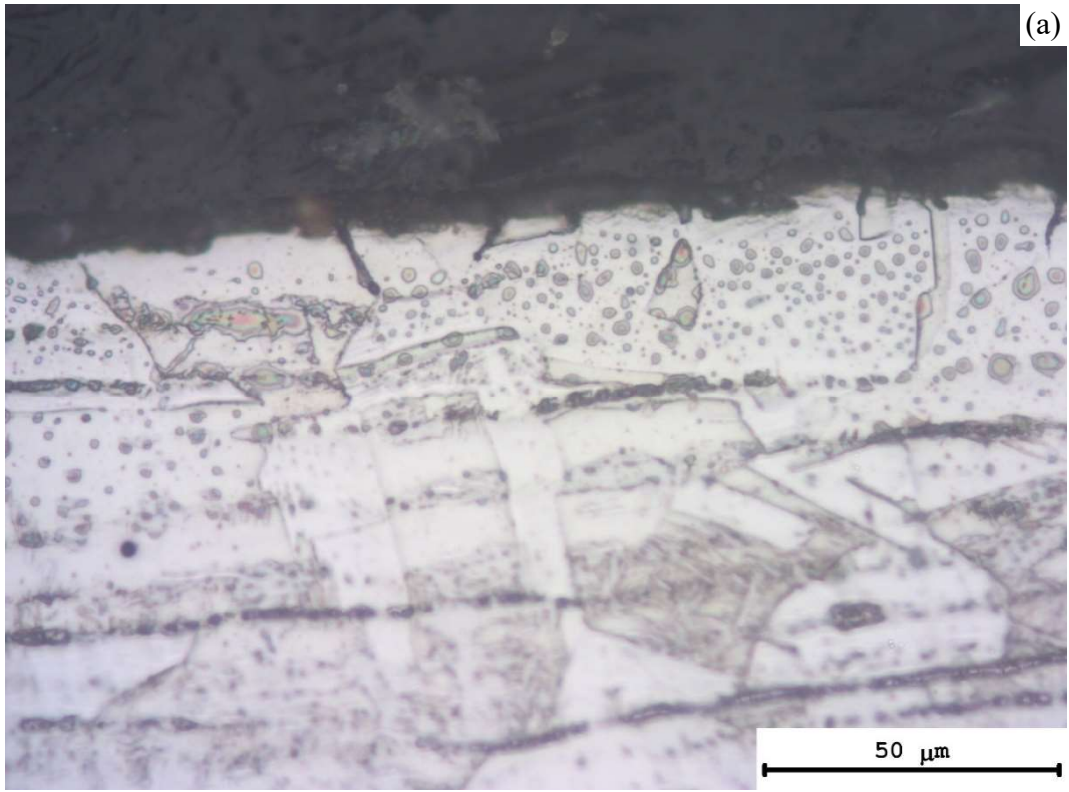

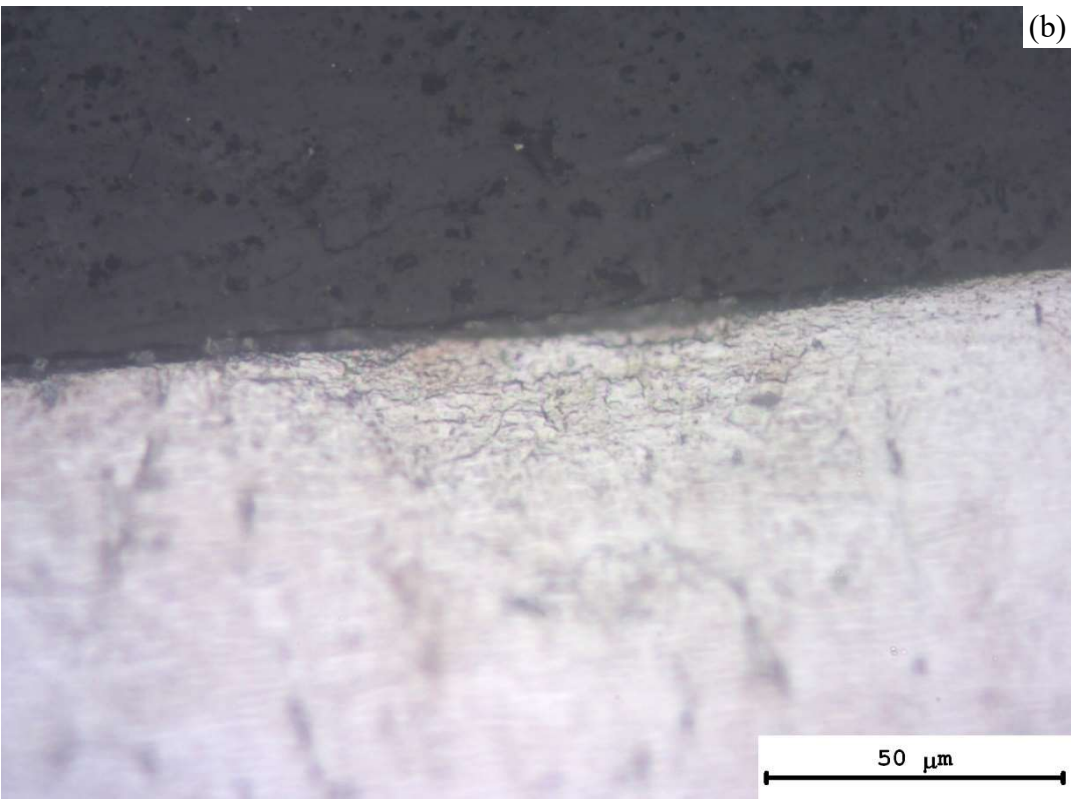

Figure 18

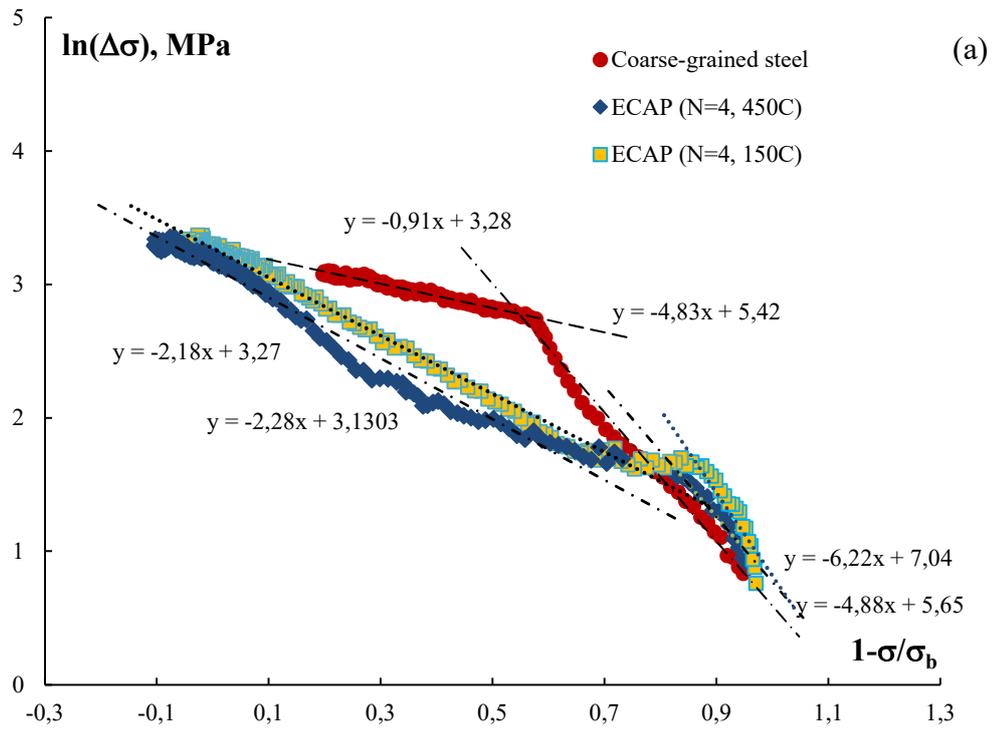

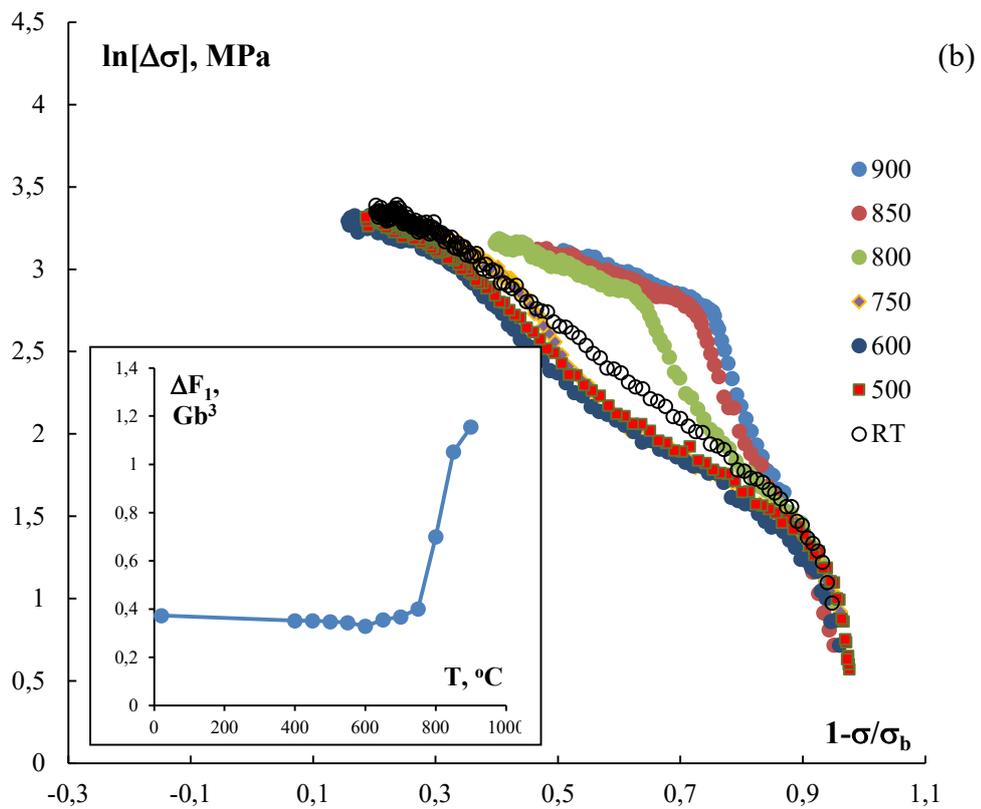

Figure 19